\documentstyle[12pt]{article}

\global\arraycolsep=1pt
\begin{document}

\hfill    HUTP-94/A040

\hfill    hepth@xxx/9411049

\hfill    November, 1994

\begin{center}
\vspace{10pt}
{\large \bf
ANOMALIES IN INSTANTON CALCULUS\\
}
\vspace{10pt}

{\sl Damiano Anselmi}

\vspace{4pt}

{\it Lyman Laboratory, Harvard University, Cambridge MA 02138,
U.S.A.}

\vspace{12pt}

{\bf Abstract}
 \end{center}

\vspace{4pt}
I develop a formalism for solving topological field theories
explicitly, in the
case
when the explicit expression of the instantons is known. I solve
topological
Yang-Mills theory with the $k=1$
Belavin {\sl et al.} instanton and topological gravity with the
Eguchi-Hanson
instanton.
It turns out that naively empty theories
are indeed nontrivial. Many unexpected interesting hidden quantities
(punctures, contact terms, nonperturbative
anomalies with or without gravity) are revealed.
Topological Yang-Mills theory with $G=SU(2)$
is not just Donaldson theory, but contains a certain {\sl link}
theory.
Indeed, local and non-local observables have the property of
{\sl marking} cycles. Moreover, from topological gravity one learns
that
an object can be considered BRST exact only if it is so all over the
moduli
space ${\cal M}$, boundary included.
Being BRST exact in any interior point of ${\cal M}$ is not
sufficient to make
an amplitude vanish.
Presumably, recursion relations and hierarchies can be found to solve
topological field theories
in four dimensions,
in particular  topological Yang-Mills theory with $G=SU(2)$ on ${\bf
R}^4$ and
topological gravity with the full set of
asymptotically locally Euclidean manifolds.

\noindent
\eject

\section{Introduction}
\label{intro}
\setcounter{equation}{0}

Not many explicit computations have been done so far
in four dimensional topological field theory,
in particular when the ghost number anomaly is greater than zero
and one needs to insert nontrivial observables to construct
nonvanishing
amplitudes.
So, it is still not enough clear
how to compute quantities and what is the
relevance of topolgical field theory for physics.
The purpose of this paper is to look
for the basic rules to make explicit
computations in the cases in which
the explicit expressions of the instantons are known.

In quantum field theory, a theory is entirely encoded in the
functional
integral
and the set of rules for manipulating it.
Consequently, it is very important to make explicit direct
computations in the very special class of quantum field theory
represented by
topological field theory,
in order to test  whether the functional integral is well-defined or
not.
Even if it were not so, something very interesting could be learned.

All the examples considered in this paper suggest that
the functional integral is well-defined in topological field theory.
Nevertheless, one learns
that sometimes it is crucial to ``follow the
instructions'' very carefully. For example,
one can appreciate the fact that certain
problems with apparent divergences
disappear only when the equivariant
cohomology requirement
is correctly understood and fulfilled.

Some topological field theories
are believed to coincide with certain previously defined mathematical
problems.
For example, this is the case of topological Yang-Mills theory in
four
dimensions,
which is believed to be equivalent to Donaldson theory \cite{witten}.
Note, however, that in physics, once the functional integral is
specified,
i.e.\
the fields, their boundary conditions, the gauge symmetry
and the gauge-fixings, then the content of the theory is also fixed.
So, explicit computations can be a useful tool for testing the
expectations.
Moreover, some theories produced by the so-called topological twist
\cite{witten}
suggest new problems not considered before in the mathematical
literature \cite{hyper}.

One can reasonably say that in theoretical physics
a problem is satisfactorily understood and solved,
once it has been reduced to an algorithmic procedure.
This is certainly one of the reasons
for the enormous success of perturbative quantum field theory.
The situation is far from being like this today in topological field
theory, or, more generically, in instanton calculus.
In this paper, I try to reduce explicit computations
in topological field theory to an algorithmic procedure.
In section \ref{formalism} I develop the formal set-up, which
is the guide-light  for the rest of the paper, devoted to
applications of the general procedure, from which
all the results follow, sometimes unexpectedly.

To begin with, an illustrative example is presented in section
\ref{t2},
namely 2D topological gravity on the torus $T^2$ with one puncture.

In section \ref{tymt} I consider 4D topological Yang-Mills theory
with $G=SU(2)$ on $M={\bf R}^4$.
In particular I solve the theory with the
$k=1$ Belavin {\it et al.} instanton {\cite{belavin}. Explicit
computations
show that the observables have the property of
{\sl marking} cycles. In other words, amplitudes like
$<{\cal O}_{\gamma_1}\cdot {\cal O}_{\gamma_2}>$
are not necessarily trivial when $\gamma_1$ and $\gamma_2$ are
trivial in the sense of the $M$-homology.
$<{\cal O}_{\gamma_1}\cdot {\cal O}_{\gamma_2}>$
can be different from zero
if $\gamma_1$ is nontrivial in the $M\backslash\gamma_2$-homology and
$\gamma_2$ is nontrivial in the $M\backslash\gamma_1$-homology.
So, topological Yang-Mills theory is not just Donaldson theory, but
contains a certain {\sl link} theory \cite{horowitz}.

In section \ref{tg} I consider 4D topological gravity with
the Eguchi-Hanson instanton. I solve the theory and compute some
{\sl anomalous} amplitudes, that turn out to be nonzero even if the
observables
are integrated
over representatives of trivial cycles. Although there
is no evidence, yet, of a link theory contained in topological
gravity,
these nonzero results are due to a subtlety related to
the boundary of
the moduli space, that is able to turn a naively BRST exact object
into a nontrivial one. In this sense, such anomalies look like
the holomorphic anomaly of ref.\ {\cite{cecotti}. Moreover, I
show that these anomalies are {\sl necessarily} present.
They reveal that an object can only be considered BRST
exact if it is so all over the moduli space, boundary included.
If it is BRST exact in any interior point of the moduli space, but
not
at the boundary, then it is not BRST exact.
In section \ref{couple} I couple topological gravity to abelian
topological
Yang-Mills theory and compute some other anomalous amplitudes.

\section{Formalism}
\label{formalism}
\setcounter{equation}{0}

In this section I develop a formalism for solving a topological field
theory
explicitly, when the explicit expression of  the instanton is known.
Just to fix notation, I manipulate the BRST algebra of topological
Yang-Mills
theory with gauge group $G$ on a manifold $M$ with instantons defined
by the
self-duality of the field strength.
Nevertheless, the method is completely general, as I show in the
applications.
In the last part of the section, I briefly discuss some
straightforward
modifications required by topological gravity.

First of all, what do we mean by {\sl solving} a topological field
theory
explicitly?
A topological field theory can be seen as a map $\pi:H(M)\times {\cal
A}\longrightarrow {\cal H}({\cal M})$ acting from the homology $H(M)$
of the
manifold $M$ times the algebra ${\cal A}$ of observables ${\cal O}$,
to the
cohomology
${\cal H}({\cal M})$ of the moduli space ${\cal M}$ of
instantons\footnotemark
\footnotetext{Actually, it is Donaldson theory that is a map
$H_k(M)\rightarrow
{\cal H}^{4-k}({\cal M})$. Commonly, Donaldson theory and topological
Yang-Mills theory
are believed to be equivalent \cite{witten}. However, one of the
byproducts of
the investigation carried on in the present paper is that it is not
so, rather
topological Yang-Mills theory contains more,
in particular a certain kind of {\sl link} theory.
Then, the definition of the map $\pi$ will have to be suitably
amended.}.
Thus, solving a topological field theory amounts to find the map
$\pi$, i.e.\
writing down the observables ${\cal O}_\gamma$, $\gamma\in H(M)$  as
forms
$\omega_\gamma({\cal O})\in {\cal H}({\cal M})$ on the moduli space
${\cal M}$.
The physical amplitudes are then the integrals over ${\cal M}$ of
top-forms
constructed with
$\omega_\gamma({\cal O})$.

If $A_\mu^a$ is the instanton, i.e.\ a {\sl coordinate} on ${\cal
M}$, let
$\delta A_\mu^a$
denote the {\sl differential} on ${\cal M}$. Then, one would like to
find the
explicit expressions for
\begin{equation}
\omega_\gamma({\cal O})={\cal F}^{({\cal O}_\gamma)}_{i_1,\ldots
i_n}(A)\delta
A^{i_1}\wedge\cdots\wedge\delta A^{i_n},
\end{equation}
for any $\gamma\in H(M)$ and any observable ${\cal O}\in{\cal A}$.

\subsection{Topological Yang-Mills theory}

\subsubsection{BRST algebra and observables}
Before seeing how this can be achieved, let us write down the BRST
algebra of
topological Yang-Mills theory in standard notation
\begin{eqnarray}
sA_\mu^a&=&\psi_\mu^a+D_\mu C^a\equiv\psi^{\prime a}_\mu,\nonumber\\
s\psi_\mu^a&=&-D_\mu\phi^a-{f^a}_{bc}\psi^b_\mu C^c,\nonumber\\
s\phi^a&=&{f^a}_{bc}\phi^bC^c,\nonumber\\
sC^a&=&\phi^a-{1\over 2}{f^a}_{bc}C^bC^c,
\label{brsym}
\end{eqnarray}
where
$D_\mu^{ab}=\delta_{ab}\partial_\mu+f^{a\phantom{c}b}_{\phantom{a}c}A_
\mu^c$
and
$f^a_{\phantom{a}bc}$ are the structure constants of $G$.

The observables are generated by the BRST extensions of identities
like
$d\,{\rm tr}\, [F\wedge F]=0$. For future use, I write down here the
simplest
ones, namely
\begin{eqnarray}
{\cal O}_x^{(0)}&=&{\rm tr}[\phi\phi](x),\quad\quad
{\cal O}_{\gamma_1}^{(1)}=2\int_{\gamma_1}{\rm
tr}[\psi\phi],\nonumber\\
{\cal O}_{\gamma_2}^{(2)}&=&\int_{\gamma_2}{\rm tr}[\psi\psi+2 F
\phi],
\quad\quad
{\cal O}_{\gamma_3}^{(3)}=2\int_{\gamma_3}{\rm tr}[F \psi],
\label{obs}
\end{eqnarray}
$\gamma_i$ being representatives of $i-$cycles on $M$. Note that the
integrands
are differential forms on $M$ and have a ghost number. Let us call
such objects
 {\sl ghost-forms}. When dealing with ghost-forms, the relevant
grading is the
{\sl ghost-form number}, i.e.\ the sum of  the ghost number and the
form
degree. The integrands in (\ref{obs}) have ghost-form number 4, form
degree $i$
and
ghost number $4-i$. The integrals, instead, have only a ghost number.
Consequently, for consistency, we have to
assign a negative form degree to the integral symbols, so that
$\int_{\gamma_i}$ has form degree $-i$. This has to be kept into
account when
commuting the BRST operator $s$ with integrals.

I am going to show that all the task for solving the theory
amounts to find the explicit expression of the ghost $C^a$.
The equation that determines it will be written down in a moment.
The concrete examples that I will discuss in the next section
suggest that whenever the explicit expression of the instanton is
known, the
equation for $C^a$  can also be solved explicitly.

\subsubsection{Gauge-fixings}
In general, the instanton $A_\mu^a$ will satisfy a certain
gauge-fixing
condition, $\partial^\mu A_\mu^a=0$ for example. The explicit form of
this
gauge-fixing condition
is totally immaterial to our purpose, as it will be clear in the
sequel. One
only needs to know that the
instanton satisfies a {\sl certain} gauge-fixing condition. Instead,
what is
crucial
is the gauge-fixing condition for the topological ghost $\psi^a_\mu$.
It is
this condition
that determines $C^a$ and solves the problem. So, I must discuss this
gauge
condition in detail.

The role of the gauge-fixing condition for $\psi_\mu^a$ is to fix the
so-called
{\sl gauge of the gauge}.
Indeed, in topological field theory one frequently has to do with a
hierarchy
of gauge-symmetries. In the case we are dealing with, it is
convenient to
distinguish
three such symmetries:
the first one is the topological symmetry (ghost  $\psi_\mu^a$) and
is the most
important;
the second one is the gauge of the gauge, i.e.\
the symmetry that acts on $\psi_\mu^a$ like an ordinary
gauge-symmetry
acts on $A_\mu^a$ (the ghost is $\phi^a$, which, to be precise, is
called ghost
for the ghosts); the third one
is the ordinary gauge-symmetry (ghost $C^a$).
These three symmetries are combined together so as to produce a
nilpotent
($s^2=0$) BRST
operator $s$.

Let us now describe the gauge-fixings.
The instantonic condition ($F_{\mu\nu}^{+a}=0$, for example) is the
gauge-fixing of the first symmetry and has to preserve the other two.
The
gauge condition we are looking for, instead, has to break the
second symmetry (and eventually also the first), while preserving the
third
one.
Finally, the usual condition $\partial^\mu A_\mu^a=0$ breaks the
third symmetry (it can also break the other two).
These are the requirements of the so-called equivariant
BRST-cohomology.
If one does not satisfy them, then one gets wrong or meaningless
results.

Instead, breaking the second symmetry while preserving the third one
is
crucial: to be more explicit, a condition
like $\partial^\mu \psi_\mu^a=0$ is wrong,
while a condition like $D^\mu \psi_\mu^a=0$ is correct. Indeed,
the condition $\partial^\mu \psi_\mu^a=0$ would kill all the local
observables.
It is obvious that $\psi_\mu^{\prime a}=\delta A^a_\mu$
satisfies $\partial^\mu \psi_\mu^{\prime a}=0$:
this is proved simply by taking the $\delta$-variation of
$\partial^\mu
A_\mu^a=0$
($s$ and $\delta$ are essentially the same, as far as our
calculations are
concerned).
Then one can take $\psi_\mu^{\prime a}= \psi_\mu^a$ and $C^a$=0.
This implies $\phi^a=0$: all the local observables
(like ${\rm tr}[\phi^2]$) vanish. So, our gauge-fixing for the gauge
of the
gauge
will be $D^\mu \psi_\mu^a=0$.

\subsubsection{Solution}
Now, we are ready to describe the procedure for solving a topological
field
theory.
$A_\mu^a$ is explicitly known by assumption.
$\psi^{\prime a}_\mu=\delta A_\mu^a$ is found by a simple
differentiation with
respect to the moduli.
Writing $\psi_\mu^a=\psi^{\prime a}_\mu-D_\mu C^a$, the condition
$D^\mu \psi_\mu^a=0$ then becomes an equation for $C^a$:
\begin{equation}
D_\mu D^\mu C^a=D^\mu \psi_\mu^{\prime a}.
\label{feq}
\end{equation}
Once this equation is solved, $C^a$ and $\psi_\mu^a$ are found.
It is clear that $\psi_\mu^a$ satisfies its field equation, whatever
$C^a$ is:
the
$\psi_\mu^a$ field equation (in our case $D_{[\mu}\psi_{\nu]^+}^a=0$)
is the $\delta$-variation of the instantonic condition
($F_{\mu\nu}^{+a}=0$)
and does not depend on $C^a$ because the instantonic condition has to
preserve
the third symmetry.

Finally, $\phi^a$ is found as
\begin{equation}
\phi^a=\delta C^a+{1\over 2}{f^a}_{bc}C^bC^c.
\end{equation}
Again, $\phi^a$ automatically satisfies its field equation.

Let $G^{ab}(x,y)$ denote the Green function for equation (\ref{feq}),
\begin{equation}
[D_\mu D^\mu(x)]^{ac}G^{cb}(x,y)=\delta^{ab}\delta (x-y).
\end{equation}
In many cases $G^{ab}(x,y)$ has been worked out explicitly
{\cite{green}.
We shall not need the explicit form of $G^{ab}(x,y)$ in our
calculations.
One can formally write
\begin{eqnarray}
C^a(x)&=&\int_M G^{ab}(x,y)D^\mu\delta A_\mu^b(y)\,  dy,\nonumber\\
\psi_\mu^a(x)&=&\int_M\{\delta^{ab}\delta(x-y)\delta_\mu^\nu-[D_\mu(x)
]^{ac}G^{cd}(x,y)
[D^\nu(y)]^{db}]\}\, \delta A_\nu^b (y) \, dy.
\end{eqnarray}
$C^a$ and $\psi_\mu^a$ have been written as one-forms on ${\cal M}$.
Writing down $\phi^a$ according to the above prescription, one finds
\begin{equation}
\phi^a(x)=\int_M G^{ab}(x,y){f^a}_{bc}\psi_\mu^b(y)\psi^{\mu c}(y) \,
dy.
\end{equation}
Indeed, taking the BRST variation of (\ref{feq}) one checks that
$\phi^a$
satisfies
\begin{equation}
D_\mu D^\mu \phi^a=f^a_{\phantom{a}bc}\psi^b_\mu \psi^c_\mu.
\label{eqfi}
\end{equation}

\subsubsection{Functional integral}
One may wonder  when the functional integral enters the game.
Since the topological field theory action is a set of gauge-fixings,
one can
simply deal
with the gauge-fixing conditions directly, as we have done.
To justify this, let us choose the gauge-fermion \cite{baulieu}
\begin{equation}
\Psi=\bar\chi_a^{\mu\nu}F^{+a}_{\mu\nu}+\bar\phi_a D^\mu\psi_\mu^a+
\bar C_a \partial^\mu A^a_\mu.
\end{equation}
Then, the Lagrangian will be the BRST variation of this gauge-fermion
(plus
eventually the topological invariant, that I do not include, since I
shall work
with a fixed instanton number):
\begin{eqnarray}
{\cal L}=s\Psi&=&b^{\mu\nu}_aF^{+a}_{\mu\nu}-\bar\chi_a^{\mu\nu}
D_{[\mu}\psi^a_{\nu]^+}+
\bar\eta_aD^\mu\psi^a_\mu-\bar\phi_a(D_\mu D^\mu\phi^a+
f^a_{\phantom{a}bc}
\psi_\mu^b\psi^{c\mu})\nonumber\\&&
+b_a\partial^\mu A^a_\mu-\bar C_a (\partial_\mu D^\mu C^a
+\partial^\mu\psi^a_\mu).
\end{eqnarray}
Integrating the Lagrange multipliers $b^{\mu\nu}_a$, $\bar\eta_a$ and
$b_a$ and
the antighosts
$\bar\chi_a^{\mu\nu}$, $\bar\phi_a$ and $\bar C_a$ away (this is
allowed, since
the observables do not depend on these fields), one gets a set of
delta
functions that agree with the equations that we wrote and solved
before.
In this way, the functional integral is performed exactly and there
is no
perturbative correction.
Instead, the Lagrangian that Witten wrote in \cite{witten},
obtained by twisting N=2 super Yang-Mills theory, contains extra BRST
exact
\cite{baulieu} terms, which spoil the linearity of ${\cal L}$ in
antighosts.
Indeed, there is no need to introduce the full set of renormalizable
interactions in the gauge-fixing sector \cite{dam,dam2}: the theory
does not
depend on (the continuous deformations of) the gauge-fixing. The role
of the
gauge-fixing sector is that of permitting to define the propagators
and a minimal choice is quite sufficient. On the other hand, one
could even
choose a power-counting non-renormalizable gauge-fixing,
without affecting the results \cite{dam}.

\subsubsection{Functional measure}
The functional measure $d\mu$ is not a problem in
topological field theory, since there exists a canonical one
\cite{witten}, which reads
\begin{equation}
d\mu=dm\, d\hat m,
\label{meas}
\end{equation}
$m$ denoting the moduli and $\hat m$ their ghost partners ($\hat
m=sm$).
To pass from the original functional measure to the above one,
one has to deal with Jacobian determinants, that, however,
mutually simplify between bosons and fermions. Then, the net effect
of
(\ref{meas}) is that of replacing, {\sl via}
the $\hat m$-integration, $\hat m$ with $dm$ wherever $\hat m$
appears.

\subsubsection{Zero modes}
I shall only consider cases in which $C^a$ and $\phi^a$ possess no
zero modes
or
their zero modes can be simply dealt with. In general, one must
include them in
the most general solution to (\ref{feq}). Getting rid of them in the
physical
amplitudes
requires the introduction of suitable {\sl puncture operators}. It is
better to
stop a moment and discuss this point in general, because it will be
useful in
the applications.

So, let us assume that $C^a$ possesses zero modes:
\begin{equation}
C^a=\tilde C^a+C^a_0,\quad\quad C^a_0(x)=C^a_i(x)\theta^i,
\end{equation}
$\tilde C^a$ denoting a particular solution to the inhomogeneous
equation
(\ref{feq}), and
$C^a_0$ denoting the most general solution to the homogeneous
equation $D_\mu
D^\mu C^a_0=0$, expanded in a basis $C^a_i(x)$. Eq.\ (\ref{eqfi})
shows that
$\phi^a$ has the same zero modes. We write
$\phi^a=\tilde\phi^a+\phi^a_0$,
$\phi^a_0(x)=\phi^a_i(x)\gamma^i$.
$\theta_i$ have ghost number $1$, while $\gamma_i$ have ghost number
$2$. So,
they correspond to 1-forms and 2-forms on ${\cal M}$, respectively,
and we have
to work out the explicit expressions of these forms.

Zero modes are to be regarded as a further symmetry of the Lagrangian
${\cal
L}$.
I call it the {\sl zero mode symmetry}.
Since ${\cal L}$ does not depend on zero modes, the functional
integral is
still ill-defined: fermionic zero modes integrate to give zero, while
bosonic
zero modes integrate to give
infinity. There is a very well-established
machinery to treat such problems, which is the BRST
technique. So, the first thing to do is to choose a gauge-fixing for
the zero
mode symmetry.
This can be the requirement that $C^a(x)$ vanishes in a certain set
of points
$\{x_i\}$.
Let us introduce antighosts $\bar\gamma_i$ and Lagrange multipliers
$\bar
\theta_i$ of ghost numbers $-2$ and $-1$, respectively
($s\bar\gamma_i=\bar\theta_i$, $s\bar\theta_i=0$).
The gauge-fermion is
\begin{equation}
\Psi_0=\sum_i\bar \gamma_i C^a(x_i).
\end{equation}
The total Lagrangian is ${\cal L}_{\rm tot}={\cal L}+{\cal L}_0$,
where
\begin{equation}
{\cal L}_0=s\Psi_0=\sum_i\bar\theta_iC^a(x_i)+\sum_i\bar
\gamma_i\left(\phi^a(x_i)+{1\over
2}f^a_{\phantom{a}bc}C^b(x_i)C^c(x_i)\right).
\end{equation}
Integrating $\bar\theta_i$ and $\bar \gamma_i$ away, one ends up with
the
insertion of the following puncture operator:
\begin{equation}
P=\prod_{i,a}C^a(x_i)\delta[\phi^a(x_i)].
\label{punct}
\end{equation}
Now, the conditions $C^a(x_i)=0$ and $\phi^a(x_i)=0$, imposed by this
insertion,
can be easily recognized as equations of $\theta_i$ and $\gamma_i$,
that
permit to find their explicit expressions as forms
on the moduli space ${\cal M}$.

\phantom{.}

Collecting the information that we have found so far,
we can conclude that the BRST algebra  is solved consistently, so
that
the map $\pi$ can be written down and the amplitudes can be
calculated. I shall do this explicitly in section \ref{tymt},
for $G=SU(2)$, $k=1$ and $M={\bf R}^4$.

\subsection{Topological gravity}

Here I briefly describe how to deal with topological gravity. This
will be used
in sections \ref{tg}
and \ref{couple}.

Several formalisms appeared in the literature for writing down the
BRST algebra
and the
observables of topological gravity (in arbitrary dimension)
\cite{literature}. The
computations of sections \ref{tg} and \ref{couple} will show
that the most convenient formalism is the most similar
to the one used for topological Yang-Mills theory.
In particular, it is important to express the observables in a simple
form.
As it was shown in sect.\ 3 of \cite{mepie} (see there for further
details),
one can write
the BRST algebra as
\begin{eqnarray}
se^a&=&\psi^a-{\cal D}\varepsilon^a-\varepsilon^{ab}e^b=\psi^{\prime
a}=
\psi^{\prime ab}e^b,\nonumber\\
s\omega^{ab}&=&\chi^{ab}-{\cal D}\varepsilon^{ab}\equiv\chi^{\prime
ab},\nonumber\\
s\psi^a&=&-{\cal
D}\phi^a-\varepsilon^{ab}\psi^b+\chi^{ab}\varepsilon^b+\eta^{ab}e^b,\nonumber\\
s\chi^{ab}&=&\chi^{ac}\varepsilon^{cb}-\varepsilon^{ac}\chi^{cb}-{\cal D}\eta^{ab},\nonumber\\
s\phi^a&=&\eta^{ab}\varepsilon^b-\varepsilon^{ab}\phi^b,\nonumber\\
s\eta^{ab}&=&\eta^{ac}\varepsilon^{cb}-\varepsilon^{ac}\eta^{cb},\nonumber\\
s\varepsilon^a&=&\phi^a-\varepsilon^{ab}\varepsilon^b,\nonumber\\
s\varepsilon^{ab}&=&\eta^{ab}-\varepsilon^{ac}\varepsilon^{cb}.
\label{brsgrav}
\end{eqnarray}
Note that there is a change in notation with respect to \cite{mepie}:
the spin
connection and curvature are defined so that
$R^{ab}=d\omega^{ab}+\omega^{ac}\omega^{cb}$. This is just to fit
with
the notation in which the Eguchi-Hanson metric is commonly written
down.

I now discuss how the procedure for solving a topological theory
has to be applied to gravity.
We have learned that it is crucial to choose correct gauge-fixing
conditions
for $\psi^{a}=\psi^{ab}e^b$. Convenient ones are
\begin{equation}
{\cal D}_b\psi^{ab}=0,\quad\quad \psi^{ab}=\psi^{ba},
\label{gfix}
\end{equation}
where ${\cal
D}_c\psi^{ab}=e^\mu_c\partial_\mu\psi^{ab}-\psi^{ae}\omega^{eb}_c+\omega^{ae}_c\psi^{eb}$.
Being $\psi^{ab}$ and $\varepsilon^{ab}$ symmetric and antisymmetric,
respectively, the relation $\psi^{\prime ab}=\psi^{ab}+{\cal
D}^b\varepsilon^a-\varepsilon^{ab}$ allows us to write
\begin{equation}
\psi^{ab}=\psi^{\prime\{ab\}}-{\cal
D}^{\{a}\varepsilon^{b\}},\quad\quad
\varepsilon^{ab}=-\psi^{\prime[ab]}-{\cal D}^{[a}\varepsilon^{b]}.
\label{gfsol}
\end{equation}
This solves the second equation of  (\ref{gfix}).
Instead, the condition ${\cal D}_b\psi^{ab}=0$ becomes a differential
equation
for $\varepsilon^a$.
Once it is solved, $\psi^{ab}$ and $\varepsilon^{ab}$ are determined.
All the
rest follows by a straightforward application of the formul\ae\
(\ref{brsgrav}). $\varepsilon^{a}$
plays the role that was played by $C^a$ in topological Yang-Mills
theory.

The observables we shall deal with are derived from the BRST
extension of
$d\, {\rm tr}[R\wedge R]=0$ ($d\, {\rm tr}[R\wedge \tilde R]=0$ will
give the
same thing),
in particular
\begin{equation}
{\cal O}_{\gamma_1}^{(1)}=2\int_{\gamma_1}{\rm tr}[\chi\eta],\quad
{\cal O}_{\gamma_2}^{(2)}=\int_{\gamma_2}{\rm tr}[\chi\chi+2 R \eta],
\quad
{\cal O}_{\gamma_3}^{(3)}=2\int_{\gamma_3}{\rm tr}[R \chi],
\label{grobs}
\end{equation}
while the local observable ${\rm tr}[\eta^2]$ is not
interesting to our problem, since we shall focus on metrics with less
than four
moduli.
In section \ref{couple} I also consider, in abelian topological
Yang-Mills
theory coupled to topological gravity, observables derived from
identities like $d\,[({\rm tr}[RR])^mF^n]=0$.

\subsubsection{Changes of variables}
Before concluding this section, I discuss what happens when  changing
coordinates.
It is useful to work on the bundle $X$ that has ${\cal M}$ as base
manifold and
$M(m)$ as fiber on $m\in{\cal M}$ (this is sometimes called the
{\sl tautological bundle}), because
the BRST extended objects (like $\hat e^a=e^a+\varepsilon^a$,
$\hat\omega^{ab}
=\omega^{ab}+\varepsilon^{ab}$, etc.; let me call them {\sl hatted
forms}) are
differential forms on $X$ and $\hat d=d+s$ is the exterior derivative
on $X$.
The points of the manifold $X$ are denoted by $(x,m)$ and we are
interested in
generic moduli-dependent changes of variables on $M$, which have the
form
$x=x(x^\prime,m)$. These changes of variables can be useful in many
applications, because sometimes it is convenient to parametrize
cycles in a
very peculiar coordinate system.
One could simply re-start from the beginning,
working out the new fields and the new BRST transformations in the
new
reference frame. However, the two solutions can be simply related to
each
other, overcoming the problem that a BRST variation (i.e. the
derivative with
respect to $m$)
at fixed $x$ is essentially different from the BRST variation at
fixed
$x^\prime$.
The answer is the following. The hatted forms are unaffected, but the
decompositions according to ghost number and form degree have to be
rederived.
Indeed,
\begin{equation}
dx=dx^\prime\left.{\partial x\over \partial
x^\prime}\right|_m+dm\left.
{\partial x\over \partial m}\right|_{x^\prime},
\end{equation}
so that forms on $M$ acquire ghost terms. For example, the vielbein
becomes the
sum of an $(1,0)$ piece (the new vielbein) plus a $(0,1)$ piece.
The latter  has to be added to $\varepsilon^a$, so that the hatted
form
$\hat e^a=e^a+\varepsilon^a$ is unaffected.

In particular, the BRST extension of the torsion is $\hat
R^a=R^a+\psi^a+\phi^a$. $\psi^a$ cannot acquire ghost terms from
$R^a$, because
we work at vanishing torsion. So, as far as the new $\psi^a$ is
concerned, one
can forget about the $m$ dependence in $x=x(x^\prime,m)$.
That means that the new $\psi^{ab}$ is related to the old one by a
diffeomorphism and a Lorentz rotation (i.e.\ by a transformation of
the {\sl third} gauge symmetry). We know that the gauge conditions
that break the second gauge symmetry have to respect the
third one. So, we are guaranteed that the new $\psi^{ab}$ satisfies
them,
if the old one did. Concretely, in our case (\ref{gfix}) are
preserved by
changes of variables.
This is another example of the importance of the equivariant
cohomology
prescription and at the same time shows that  notions like
differential forms
on
$M$ and ghost fields have not an {\sl invariant} meaning. Only hatted
forms
have an invariant meaning.

\subsection{Conclusion}

The procedure elaborated so far works for many cases and I will not
go
further in this paper. Thus, I shall also be able to compute
well-defined
amplitudes
in power-counting non-renormalizable quantum field theories (like
quantum
gravity), by
studying their topological versions (which are perturbatively finite:
there is
no beta
function, because there is no coupling constant).
As discussed in \cite{dam}, the infinitely many
types of counterterms that may appear perturbatively do not affect
the
topological results
(in absence of anomalies). If, instead, there are anomalies, then the
theory
could be even more interesting, since the breaking of the topological
symmetry
could generate quantum gravity.
In perturbation theory, an anomaly could turn some gauge-fixing
parameter into
a physical one. Then, one should
count the number of such anomalies: if they are a finite number, they
generate
a predictive theory. This motivation is quite sufficient to justify
such a kind
of investigation.

The following sections are devoted to applications of the formalism
established
in the present one. I begin with an illustrative very simple case and
then I
turn to topological Yang-Mills theory and topological gravity.

\section{A trivial example: $T^2$}
\label{t2}

As a first example, I consider a very simple case that leads to a
non-vanishing
amplitude:
the torus $T^2$ in two dimensional topological gravity. It will be
described by
the cube
$(\xi,\eta)\in[0,1]\times [0,1]$ with flat metric
\begin{equation}
ds^2=d\xi^2+|\tau|^2d\eta^2+2\,  {\rm Re}\,\tau \,\,d\xi d\eta.
\end{equation}
Coordinates like $z=\xi+\tau \eta$ are not good for the computational
procedure
that is
described in this paper. In a field theoretical approach, the
coordinates must
parametrize the topological manifold and the moduli are fields.

I choose the zweibein
\begin{equation}
e^0=d\xi+{\rm Re}\,\tau \,\,d\eta,\quad\quad e^1={\rm Im}\,\tau\,\,
d\eta.
\end{equation}
The variation of the zweibein is
\begin{equation}
se^a=\psi^a-{\cal D}\varepsilon^a-\varepsilon^{ab}e_b=\psi^{\prime
ab}e_b,
\end{equation}
$\varepsilon^a$ and $\varepsilon^{ab}$ being the diffeomorphism and
Lorentz
rotation ghosts, respectively.
One finds
\begin{equation}
\psi^{\prime ab}=\left(\matrix{0&i{d\tau+d\bar\tau\over
\tau-\bar\tau}\cr
0&{d\tau-d\bar\tau\over \tau-\bar\tau}}\right).
\end{equation}
Imposing the gauge-conditions (\ref{gfix}) on $\psi^a=\psi^{ab}e_b$,
it turns out that $\varepsilon^a$ is constant, while $\psi^{ab}$ and
$\varepsilon^{ab}$ are the symmetric and antisymmetric parts of
$\psi^{\prime
ab}$, respectively. In particular,
\begin{equation}
\varepsilon^{ab}=-{i\over 2}{d\tau+d\bar\tau\over
\tau-\bar\tau}\left(\matrix{0&1\cr -1&0}\right)
\equiv c_0\left(\matrix{0&1\cr -1&0}\right).
\end{equation}
The local observable $\gamma_0$ of the theory is the BRST
variation of $c_0$, i.e.\
\begin{equation}
\gamma_0=i{d\tau\wedge d\bar \tau\over (\tau-\bar \tau)^2},
\end{equation}
which is the Poincar\`e metric, the Poincar\`e dual of a point. Thus
the
amplitude $<\gamma_0>$ is the volume of the moduli space of the
torus, which is
certainly a well-defined amplitude.
In the usual description \cite{verlindesquare}, it corresponds to the
amplitude
$<\sigma_1(x)>$.
The presence of one puncture is revealed by the fact that
$\varepsilon^a$ has
two real zero modes (the constants) and so one has to introduce a
puncture
operator to get rid of them.

Knowing the explicit expression for the metric of the torus with more
punctures
would allow to recover the remaining amplitudes.
It could be interesting to recover the full set of amplitudes
of the punctured sphere, first.
Perhaps, combining the formalism developed in the previous section
with Strebel's theory of quadratic differentials on Riemann surfaces
\cite{strebel} one can recover Kontsevich's result \cite{kontsevich}.

\section{Topological Yang-Mills theory}
\label{tymt}
\setcounter{equation}{0}

In this section, I consider topological Yang-Mills theory on $M={\bf
R}^4$ with
gauge group $G=SU(2)$.  Let $k$ denote the instanton number.
Then the moduli space ${\cal M}$ has dimension $8k-3$.
$H(M)$ does not contain nontrivial cycles other than the point
and $M$ itself. So, in eq.\  (\ref{obs}) only the observable
${\rm tr}[\phi^2]$ is nontrivial and the selection rule can never be
fulfilled.

I shall answer the question: is this theory empty? There is surely a
non-empty
intersection theory on the instanton
moduli space ${\cal M}$ and the topological field theory should
contain
at least a part of it. Actually, we shall see that the theory
contains many unexpected things, more related to the manifold $M$
than
to intersection theory on the moduli space ${\cal M}$.

I focus on the well-known $k=1$ instanton \cite{belavin}
\begin{equation}
A_\mu^a(x)={2\over D}\eta^a_{\mu\nu}(x-x_0)^\nu, \quad \quad
D=(x-x_0)^2+\rho^2.
\end{equation}
Here, $x_0\in {\bf R}^4$ and $\rho\in (0,\infty)$ are the moduli, so
that
${\cal M}=(0,\infty)\otimes {\bf R}^4$.
In the sequel, I strictly follow the notation of \cite{thooft}.
I do not introduce any compactification of $M$ or ${\cal M}$:
it is not necessary in my approach, because, since everything
follows automatically from the physical formalism
(functional integral manipulated as explained in section
\ref{formalism}), that will produce well-defined finite results,
one can say that, in some sense, a privileged kind of
compactification is already encoded in it.
Anyway, one can consider $M={\bf R}^4$ as a chart of $S^4$.
The point at infinity will be treated appropriately in a moment.

According to (\ref{brsym}), one has
\begin{equation}
\psi_\mu^{\prime a}=\delta A_\mu^a=-{2\over D^2}\,
\eta^a_{\mu\nu}[Ddx_0^\nu+2
(x-x_0)^\nu(\rho d\rho-(x-x_0)\cdot dx_0)],
\end{equation}
so that equation (\ref{feq}) becomes
\begin{equation}
D_\mu D^\mu C^a=-{8\over D^2}\eta^a_{\mu\nu}(x-x_0)^\mu dx_0^\nu.
\label{4.3}
\end{equation}
Note that the right hand side does not contain $d\rho$. Thus, $C^a$
has the
form
$C^a=g^a_\mu dx_0^\mu$. A natural ansatz is
\begin{equation}
C^a=f(D)\eta^a_{\mu\nu}(x-x_0)^\mu dx_0^\nu.
\end{equation}
(\ref{4.3}) reduces to the following equation for $f$:
\begin{equation}
f^{\prime\prime}(D-\rho^2)+3f^\prime+2{\rho^2\over D^2}f+{2\over
D^2}=0,
\end{equation}
the primes denoting derivatives with respect to $D$. This equation is
solved by
$f={2\over D}$, so that
\begin{equation}
C^a={2\over D}\eta^a_{\mu\nu}(x-x_0)^\mu dx_0^\nu.
\label{ghost}
\end{equation}
Note the $\sim {1\over x}$ behavior of $A_\mu^a$ and $C^a$ for
$x\rightarrow\infty$:
it will produce well-defined topological amplitudes. In general,
the ``unphysical'' quantities like $A_\mu^a$ and $C^a$  should be
regular everywhere and bounded at infinity.
In the case of topological gravity the diffeomorphism ghosts
$\varepsilon^a$
will possess a similar $\sim {1\over x}$ behavior, while the Lorentz
ghosts
$\varepsilon^{ab}$ will tend to constants (global Lorentz rotation at
infinity).

If there are zero modes, then (\ref{ghost}) is not the most general
solution.
As a matter of fact,
in \cite{thooft} it is shown that there are the following three zero
modes,
corresponding to the isospin infinitesimal rotations,
\begin{equation}
C^a_b\theta^b={1\over
D}\eta^a_{\mu\nu}\bar\eta^b_{\mu\rho}(x-x_0)^\nu
(x-x_0)^\rho\theta^b.
\label{zerom}
\end{equation}
We see that these expressions have not a well-defined limit at
infinity, rather
(so to speak)
they have a three-parameter limit at infinity,
while (\ref{ghost}) tends to zero. One can get rid of (\ref{zerom})
following
the procedure
explained in section \ref{formalism}. We have to choose three points
$x_1$,
$x_2$ and $x_3$ where $C^a$ should vanish. The most convenient choice
seems to
be $x_1=x_2=x_3=\infty$ (more precisely, one should  choose an
arbitrary
triplet of points and then let these points tend to infinity). The
fact that
(\ref{zerom}) have a three-parameter limit at infinity says that
precisely
three conditions are necessary to make the ghost vanish there.
The expressions for $\theta^a$ as forms on the moduli space are then
very
simple:
$\theta^a=0$ $\forall a$. Consequently, (\ref{ghost}) is the correct
expression
we have to deal with. The same argument, when applied to $\phi^a$,
gives
$\gamma^a=0$ $\forall a$.

With the solution (\ref{ghost}), one can write down the explicit
expression of
any quantity.
Note the very simple expression of the BRST extension $\hat A^a$:
\begin{equation}
\hat A^a=A^a+C^a={2\over D}\, d(x-x_0)^\mu\eta^a_{\mu\nu}(x-x_0)^\nu.
\end{equation}
For $\phi^a$ one gets
\begin{equation}
\phi^a=-{2\rho\over D^2}\, \eta^a_{\mu\nu}[\rho dx_0^\mu+2(x-x_0)^\mu
d\rho)]dx_0^\nu,
\end{equation}
while $\psi^a_\mu$ turns out to be
\begin{equation}
\psi^a_\mu=-4 {\rho\over D^2}\eta^a_{\mu \nu}[ \rho
dx_0^\nu+(x-x_0)^\nu
d\rho].
\end{equation}
The BRST extension $\hat F^a$ of the field strength
$F^a=dA^a+{1\over 2}\varepsilon^a_{\phantom{a}bc}A^bA^c$ also has a
very simple
expression:
\begin{equation}
\hat F^a=F^a+\psi^a+\phi^a=-2{\rho\over
D^2}d(x-x_0)^\mu\eta^a_{\mu\nu}
[\rho d(x-x_0)^\nu-2 (x-x_0)^\nu \, d\rho].
\label{hatf}
\end{equation}
$\hat F^a$ tends to zero on $\partial {\cal M}$ sufficiently rapidly
to
assure that the amplitudes are topological.

Other useful formul\ae\ are
\begin{eqnarray}
\hat F^a\hat F^a&=&4{\rho^3\over D^4}[\rho dV(x-x_0)-4d\rho \wedge
(x-x_0)^\mu
d\sigma_\mu(x-x_0)]=\hat d \hat C,\nonumber\\
\hat C&=&\hat A^a\hat F^a-{1\over 6}\varepsilon_{abc}\hat A^a
\hat A^b \hat A^c={4\over 3}{1\over D^3}[3\rho^2+(x-x_0)^2]
(x-x_0)^\mu d\sigma_\mu(x-x_0),
\label{hal}
\end{eqnarray}
where $\hat d=d+s$.

Any expression involving $\hat A$ and not just $\hat F$ holds only
locally.
A form on ${\cal M}$ is exact only if it can be written as
$s\Lambda$,
$\Lambda$ being
constructed with the components of $\hat F$, but independent of $\hat
A$. This
is the analogue of the fact that
${\rm tr}[F\wedge F]$ is not globally exact over $M$, because it
cannot be
written as $dC$ with a gauge-invariant $C$.

\subsection{Amplitudes}

Let us now come to the observables. The full set (\ref{obs}) is
generated by
the identity
$\hat d\, {\rm tr}[\hat F\hat F]=0$.
The local observable ${\cal O}^{(0)}_x$ can be promptly written down
as a
4-form on ${\cal M}$:
\begin{equation}
\omega^{(4)}_x=\phi^a\phi^a(x)=4{\rho^3\over D^4}[\rho dV(x_0)+4
d\rho\wedge
(x-x_0)_\mu d\sigma^\mu(x_0)],
\end{equation}
where
\begin{equation}
dV(x_0)=\varepsilon_{\mu\nu\rho\sigma}dx_0^\mu dx_0^\nu dx_0^\rho
dx_0^\sigma,\quad\quad
d\sigma^\mu(x_0)=\varepsilon_{\mu\nu\rho\sigma}dx_0^\nu dx_0^\rho
dx_0^\sigma.
\end{equation}
One can explicitly check that $\omega^{(4)}_x$ is closed. By
integrating it on
a 4-cycle
in ${\cal M}$, like $\{\rho_0\}\otimes {\bf R}^4=\gamma_4$ for
example, one can
show that it is not exact:
\begin{equation}
\int_{\gamma_4}\omega^{(4)}_x=16\pi^2.
\label{o1}
\end{equation}
Let us consider another 4-cycle, $\gamma^\prime_4=(0,\infty)\otimes
S^3_r$, $r$
being the radius of a 3-sphere $S^3\subset {\bf R}^4$ centered in the
origin.
Then
\begin{equation}
\int_{\gamma_4^\prime}\omega^{(4)}_x=-384 \pi
\int_0^\infty d\rho\int_0^\pi d\theta {\rho^3 r^3
(r-x \cos \theta)\sin^2\theta \over (\rho^2+r^2+x^2-2rx\cos
\theta)^4}.
\end{equation}
Before computing the above integral, let us see what is the expected
result.
First of all, it should not depend on the representative of the
4-cycle.
Second, it should not depend on the
point $x$ that defines the 4-form $\omega^{(4)}_x$. These are nothing
but the
requirements that the amplitude be topological.
If these expectations are correct, we can choose the most convenient
values of
$r$ and $x$.
So, let us choose $r=0$, i.e.\ let us shrink the 3-sphere to a point.
Since
there is no singularity
at $r=0$, the result is zero. Independence of $r$ suggests that  the
above
integral is identically zero. However, this is not true.
Let us perform the $\rho$-integration:
\begin{equation}
\int_{\gamma_4^\prime}\omega^{(4)}_x=-32 \pi\int_0^\pi d\theta {r^3
(r-x \cos
\theta)\sin^2\theta\over (r^2+x^2-2rx\cos \theta)^2}.
\end{equation}
Now, for $r\rightarrow 0$ the above expression tends to zero, in
agreement with
the above argument, but for $r\rightarrow \infty$ it
tends to $-16 \pi^2$ and for $r=x$ it is equal to $-8 \pi^2$.
So, the amplitude is surely $r$ and $x$ dependent and does not seem
to be topological. To solve the puzzle, let us
perform the complete integration:
\begin{equation}
\int_{\gamma_4^\prime}\omega^{(4)}_x=-8 \pi^2
\left(1+{\sqrt{(r^2-x^2)^2}\over
r^2-x^2}
\right)=-8 \pi^2 (1+H(r-x)),
\label{int1}
\end{equation}
$H(x)$ denoting the step function, $H(x)=1$ for $x>0$ and $H(x)=-1$
for $x<0$.
Thus, a very simple interpretation of the result comes to one's mind:
the
amplitude ``feels''
the location $x$ of the observable or, vice versa, the local
observable
modifies the
geometry. When $r>x$ one cannot shrink the 3-sphere to zero safely:
it is
necessary to
cross the observable. On the other hand, when $r<0$ no problem is
encountered
and the result
is zero. The major consequence is the following: with the insertion
of one
local observable, there
are noncontractible 3-cycles even in ${\bf R}^4$. Keeping this in
mind, in a
moment we shall go back to our original question
(``is the theory empty?'') and we shall be able to give a negative
answer.

What I have described is the first evidence of the main subject of
the present
paper:
anomalies in instanton calculus.
As a matter of fact, I did not discover them in this example, rather
in the
case of
topological gravity with the Eguchi-Hanson metric (see the next
section).
Nevertheless,
I preferred to start with an example that everybody is more familiar
with.

3-cycles play a role in the definition of the observable ${\cal
O}^{(3)}_{\gamma_3}$
of (\ref{obs}). So, let us compute $\omega^{(1)}_r={\cal
O}^{(3)}_{S^3_r}$. It
is convenient to
use (\ref{hal}) in order to
write ${\cal O}^{(3)}_{S^3_r}$ as
\begin{equation}
{\cal
O}^{(3)}_{S^3_r}=-s\int_{S^3_r}\varepsilon_{\mu\nu\rho\sigma}C^\mu
dx^\nu
dx^\rho dx^\sigma,
\label{ouno}
\end{equation}
where $C^\mu$ is the Chern-Simons form
\begin{equation}
C^\mu={1\over 6}\varepsilon^{\mu\nu\rho\sigma}\left(
A_\nu^a\partial_\rho A_\sigma^a+{1\over 3}\varepsilon_{abc}A_\nu^a
A_\rho^b
A_\sigma^c\right)
={4\over 3 D^3} (x-x_0)^\mu[3 \rho^2+ (x-x_0)^2].
\label{417}
\end{equation}
In (\ref{ouno}), the minus sign is due to having commuted $s$ with
$\int_{S^3_r}$.
Actually, (\ref{hal}) permits to write ${\cal O}^{(3)}_{S^3_r}$ as
the above
expression plus something of the form $\int_{S^3_r}d[\cdots]$. This
extra term
vanishes, since although the $[\cdots]$ is not globally defined on
$M$,
certainly it is on $S^3_r$.

It is thus simple to prove that
\begin{equation}
\omega^{(1)}_r=df,\quad\quad f(\rho,x_0)=32\pi r^3\int_0^\pi
d\theta{(r-x_0\cos
\theta)
(3\rho^2+r^2+x_0^2-2r x_0 \cos \theta)\sin^2\theta\over
(\rho^2+r^2+x_0^2-2 r x_0 \cos\theta)^3}.
\label{ji}
\end{equation}
So, the interesting physical amplitude is
\begin{equation}
{\cal A}=<{\cal O}^{(3)}_{S^3_r}\cdot {\cal O}^{(0)}_x>=\int_{\cal M}
\omega^{(1)}_r\wedge \omega^{(4)}_x.
\end{equation}
Using (\ref{ji}) and $d\omega^{(4)}_x=0$, we get
\begin{equation}
{\cal A}=\int_{\partial {\cal M}}f\omega^{(4)}_x.
\end{equation}
Now, the boundary of the moduli space is $\partial {\cal M}=
\partial_1{\cal M}\cup \partial_2{\cal M}
\cup \partial_3{\cal M}$, where
$\partial_1{\cal M}=\{0\}\otimes {\bf R}^4$,
$\partial_2{\cal M}=\{\infty\}\otimes {\bf R}^4$ and
$\partial_3{\cal M}=(0,\infty)\otimes S^3_\infty$.
We have
\begin{equation}
\lim_{\rho\rightarrow\infty}f=0,\quad
\lim_{x_0\rightarrow\infty}f=0,\quad
\lim_{\rho\rightarrow 0}f=8 \pi^2 (1+H(r-x_0)),
\label{bound}
\end{equation}
the last limit being easily evaluated, since the integral in
(\ref{ji}) reduces
to the one
in (\ref{int1}) when $\rho\rightarrow 0$. One can easily check that
$\int_{\partial_2{\cal M}}\omega^{(4)}_x$ and $\int_{\partial_3{\cal
M}}\omega^{(4)}_x$
are finite\footnotemark\footnotetext{Such integrals are
nothing but peculiar cases of (\ref{o1}) and (\ref{int1}).},
so that ${\cal A}$ receives contribution only from $\partial_1{\cal
M}$.
We thus have
\begin{equation}
{\cal A}=64\cdot 96 \pi^3\lim_{\rho\rightarrow 0}
\rho^4\int_0^rx_0^3dx_0\int_0^\pi{\sin^2\theta\,\, d\theta\over
(x^2+x_0^2+\rho^2-2 x x_0
\cos \theta )^4}.
\end{equation}
Now, let us discuss this formula. The $\rho^4$ factor kills the full
expression
at least when
the remaining integral is regular. This surely happens for $r<x$.
Thus
\begin{equation}
{\cal A}=0,\quad\quad {\rm for}\, r<x.
\end{equation}
However, one must be careful when $r>x$, because the integral is
singular at
$x=x_0$ and
$\theta=0$ : it is better to keep $\rho$ different from zero and
take the limit at the very end. Performing the $\theta$-integration,
one gets
\begin{equation}
{\cal A}=32\cdot 48 \pi^4\lim_{\rho\rightarrow 0}\int_{\rho-{x^2\over
\rho}}^{\rho+{r^2-x^2\over \rho}}
{(v\rho+2 x^2)(v\rho-\rho^2+x^2)dv\over (v^2+4 x^2)^{5/2}},
\end{equation}
where $v=\rho+{x_0^2-x^2\over\rho}$.
Amazingly, the $v$-integration gives back the by now familiar step
function:
\begin{equation}
{\cal A}=128 \pi^4 (1+H(r-x)),
\label{conf}
\end{equation}
confirming once for all the presence of an anomaly.
Thus, there is no doubt that one should accept the fact that the
presence of
local observables
alters the theory in a visible way.

This can can be
thought as something similar to the contact terms that appear in 2D
topological
gravity
\cite{verlindesquare} and to the holomorphic anomaly of
\cite{cecotti}. Indeed, ${\cal O}^{(3)}_{S^3_r}$
is naively BRST-exact, since $S^3_r$ is naively a boundary. Thus
$<{\cal
O}^{(3)}_{S^3_r}\cdot {\cal O}^{(0)}_x>$ is expected to be zero,
since it is
the average value of a BRST exact object.
However, the naive expectation is affected by the boundary of the
moduli space:
this is revealed by formula (\ref{bound}), which clearly stresses
that  the
whole result is due to the
``instanton of zero size $\rho$''.

Summarizing, the above result can be understood as follows. When the
local
observable ${\cal O}^{(0)}_x$ is placed {\sl inside} $S^3_r$, then
$S^3_r$ is
not homotopically trivial: it cannot be contracted to a point without
crossing
either the observable placed at $x$ or the puncture (\ref{punct})
placed at
infinity. In this case, the amplitude is different from zero.
Instead, if
${\cal O}^{(0)}_x$ is placed {\sl outside} $S^3_r$, then $S^3_r$ is
homotopically trivial and the amplitude vanishes.

{}From what we have discovered so far (that is not the full story),
we can say
that,
instead of classifying the homology of $M$,
one should classify the homology of $M\backslash\{x_1,\ldots x_n\}$,
$x_1,\ldots x_n$
being the positions of the local observables and of the punctures
(\ref{punct}).
Thus, we learn that concepts like puncture, contact term and
nonperturbative
BRST
anomaly\footnotemark\footnotetext{It is worth stressing that such
BRST
anomalies do not affect the unitarity of the theory. As a matter of
fact, if
things are correctly understood, they are not anomalies at all. More
details on
this will be given in the next section.}
are meaningful even in four dimensions and without gravity. The
existence of a
correspondence (topological twist) relating
topological Yang-Mills theory to N=2 supersymmetric Yang-Mills theory
\cite{witten}
suggests that similar things should be present in ordinary
supersymmetric
theories.
This should be true also for N=1 theories, since they also possess
topological
amplitudes
\cite{report}.

We can thus conclude that the theory is all but empty:
anomalies are responsible for giving sense to
a naively empty theory.

The next question that comes naturally to one's mind is the
following: if a
local observable is able to mark a point, can a non-local observable
mark a
higher dimensional cycle? It happens that the answer is positive, so
we are
going to
discover that topological Yang-Mills theory contains a certain
{\sl link}-theory .

Consider the observables ${\cal O}^{(1)}_{\gamma_1}$ and ${\cal
O}^{(2)}_{\gamma_2}$ of
formula (\ref{obs}). They correspond to a 3-form $\omega^{(3)}$ and a
2-form
$\omega^{(2)}$ on ${\cal M}$, respectively. So, their product can be
integrated
over ${\cal M}$
to give an amplitude. To define these observables and the
corresponding ${\cal
M}$-forms,
one needs a 1-cycle and a 2-cycle on $M={\bf R}^4$. There are no
nontrivial
such cycles on
${\bf R}^4$, however, enlightened by the previous results,
we can easily figure out what happens. Let us choose $\gamma_1$
and $\gamma_2$ in such a way that $\gamma_2$ is a nontrivial
2-cycle of $M\backslash\gamma_1$ and viceversa: this should produce a
finite non-vanishing amplitude ${\cal A}$.
On the other hand, if we choose $\gamma_1$ to be a trivial loop of
$M\backslash
\gamma_2$,
we expect ${\cal A}$ to be zero. We are now going to show that it is
precisely
so.

We choose $\gamma_1$ and $\gamma_2$ in the following way. Let us
write
$M={\bf R}^4={\bf R}\otimes {\bf R}^3$, ${\bf R}$ being ``time'' and
${\bf
R}^3$ being three-space.
Then, let
\begin{equation}
\gamma_1=(-\infty,\infty)\otimes \{{\bf x}_1\},\quad \quad
\gamma_2=\{t_2\}\otimes S^2_r.
\end{equation}

One finds
\begin{equation}
{\cal O}^{(1)}_{\gamma_1}=\omega^{(3)}_{{\bf
x}_1}=d\Omega^{(2)}_{{\bf x}_1},
\quad\quad
\Omega^{(2)}_{{\bf
x}_1}={\pi}{(x_1-x_0)^i\varepsilon_{ijk}dx_0^jdx_0^k\over
[\rho^2+({\bf  x}_1-{\bf x}_0)^2]^{5/2}}[5\rho^2+2 ({\bf x}_1-{\bf
x}_0)^2].
\end{equation}
One has
\begin{equation}
{\cal A}=<{\cal O}^{(1)}_{\gamma_1}\cdot {\cal
O}^{(2)}_{\gamma_2}>=\int_{\cal
M}
\omega^{(3)}_{{\bf x}_1}\wedge \omega^{(2)}_{r}=
\int_{\partial {\cal M}}\Omega^{(2)}_{{\bf x}_1}\wedge
\omega^{(2)}_{r}.
\end{equation}
It is convenient to write $x_0$ as $(t_0,{\bf x}_0)$. Since,
$\omega^{(3)}_{{\bf x}_1}$ does not contain $dt_0$, one can focus on
the terms
of $\omega^{(2)}_r$ that contain it.
One finds
\begin{equation}
\omega^{(2)}_{r}=48 \rho^3 dt_0\int_{S^2_r}{\rho dx_0^i+d\rho\,
(x-x_0)^i\over
(t_0^2+\rho^2+ ({\bf x}-{\bf x}_0)^2)^4}\varepsilon_{ijk}dx^jdx^k.
\end{equation}
Due to the fact that
\begin{eqnarray}
\Omega^{(2)}_{{\bf x}_1}&\sim& {1\over
\rho^3}(x_1-x_0)^i\varepsilon_{ijk}dx_0^jdx_0^k,
\quad\quad {\rm for}\,\, \rho\rightarrow\infty,\nonumber\\
\Omega^{(2)}_{{\bf x}_1}&\sim& {1\over
x_0^3}(x_1-x_0)^i\varepsilon_{ijk}dx_0^jdx_0^k,
\quad\quad {\rm for}\,\,  x_0\rightarrow\infty,\nonumber\\
\lim_{\rho\rightarrow 0}\Omega^{(2)}_{{\bf x}_1}&=&
2\pi{(x_1-x_0)^i\varepsilon_{ijk}dx_0^jdx_0^k\over |{\bf x}_1-{\bf
x}_0|^3},
\end{eqnarray}
one can check that $\partial_2{\cal M}$ and $\partial_3{\cal M}$ do
not
contribute. After a straightforward manipulation, one can write
\begin{equation}
{\cal A}=32\pi\lim_{\rho\rightarrow
0}\rho^4\int_{-\infty}^{\infty}dt_0
\int_{{\bf R}^3}{\varepsilon_{ijk}dx_0^idx_0^jdx_0^k\over x_0^3}
\int_{S^2_r}{x_0^m\varepsilon_{mnp}dx^ndx^p\over (t_0^2+\rho^2+({\bf
x}-{\bf
x}_1-{\bf x}_0)^2)^4}.
\end{equation}
Let us now study this amplitude in two limiting cases,
representing the two possible situations described above.

1) Let $x_1=0$. Then, whatever $r$ is, $S^2_r$ is a nontrivial
cycle of $M\backslash \gamma_1$ and we expect a nonzero amplitude.
Indeed, one finds, after rescaling all quantities by $\rho$,
\begin{equation}
{\cal A}=128\pi^2\lim_{r\rightarrow\infty}
r^2\int_{-\infty}^{\infty}dt_0\int_{{\bf
R}^3}{\varepsilon_{ijk}dx_0^idx_0^jdx_0^k\over x_0^3}
\int_{-1}^1{udu\over (1+t_0^2+r^2+x_0^2-2r x_0 u)^4}.
\end{equation}
Integrating over $u$ and setting $s_0={1\over
2r}(x_0^2+1+t_0^2-r^2)$, one
arrives at
\begin{equation}
{\cal A}=512\pi^3\lim_{r\rightarrow\infty}{1\over r}
\int_{-\infty}^\infty
dt_0\int^\infty_{{1+t_0^2\over 2r}-{r\over 2}}{(s_0+r)ds_0\over
(s_0^2+t_0^2+1)^3}.
\end{equation}
Now  we can take the limit $r\rightarrow\infty$ and find
\begin{equation}
{\cal A}=256 \pi^4.
\label{reso}
\end{equation}

2) Let $r\rightarrow 0$, $x_1\neq 0$. Then, $\gamma_1$ is a
contractible loop
of $M\backslash \gamma_2$ and we predict ${\cal A}=0$. Indeed,
\begin{equation}
{\cal A}\sim \lim_{\rho\rightarrow
0}\rho^4\int_{-\infty}^{\infty}dt_0
\int_{{\bf R}^3}{\varepsilon_{ijk}dx_0^idx_0^jdx_0^k\over x_0^3}
{x_0^m\over  (t_0^2+\rho^2+({\bf x}_1+{\bf
x}_0)^2)^4}\lim_{r\rightarrow 0}
\int_{S^2_r}{\varepsilon_{mnp}dx^n dx^p}.
\end{equation}
Now, the factor with the limit $r\rightarrow 0$ tends to zero and the
remaining
integral is easily shown to be convergent, with a manipulation
similar to the
one of point 1). So, ${\cal A}=0$.

The final expression of the amplitude is thus
\begin{equation}
{\cal A}=128 \pi^4 (1+H(r-x_1)).
\label{conf2}
\end{equation}

We now check the above result by parametrizing the cycles in a
different way.
Let us write
$M={\bf R}^2\otimes {\bf R}^{2\, \prime}$, $x=({\bf x},{\bf
x}^\prime)$,
$x_0=({\bf x}_0,{\bf x}_0^\prime)$ and
\begin{equation}
\gamma_1=C_r\otimes \{{\bf x}_1^\prime\},\quad\quad
\gamma_2=\{{\bf x}_2\}\otimes {\bf R}^{2\, \prime},
\end{equation}
$C_r$ denoting a circle of radius $r$ centered in the origin. One
finds
\begin{equation}
\omega^{(2)}_{{\bf x}_2}=d\Omega^{(1)}_{{\bf x}_2},\quad
\Omega^{(1)}_{{\bf x}_2}=-8\pi{2\rho^2+({\bf x}_2-{\bf x}_0)^2\over
(\rho^2+({\bf x}_2-{\bf
x}_0)^2)^2}[(x_2^0-x_0^0)dx_0^1-(x_1^1-x_0^1)dx_0^0].
\end{equation}
we have
\begin{equation}
{\cal A}=\int_{\cal M}\omega^{(2)}_{{\bf x}_2}\wedge
\omega^{(3)}_r=\int_{\partial{\cal M}}
\Omega^{(1)}_{{\bf x}_2}\wedge \omega^{(3)}_r=-\lim_{\rho\rightarrow
0}\int_{{\bf R}^4}
\Omega^{(1)}_{{\bf x}_2}\wedge \omega^{(3)}_r.
\end{equation}
As before, one can check that $\partial_2{\cal M}$ and $\partial_3
{\cal M}$ do
not contribute.
One finds
\begin{equation}
{\cal A}=256 \pi^2\lim_{\rho\rightarrow 0}\rho^4\int_{{\bf
R}^2}{dx_0^0dx_0^1\over
x_0^2}\int_{\gamma_1}{x_0^0dx_1^1-x_0^1dx_1^0\over
[\rho^2+({\bf x}_0+{\bf x}_2-{\bf x}_1)^2]^3}.
\end{equation}
In the limit $r=|{\bf x}_1|\rightarrow 0$, ${\bf x}_2\neq 0$, one
finds $0$,
while for ${\bf x}_2=0$ one can go on as in the previous calculation
and find
\begin{equation}
{\cal A}=256\pi^4,
\end{equation}
in agreement with (\ref{reso}).
It is clear that choosing the cycles as
\begin{equation}
\gamma_1=\cup_i(-1)^{\sigma_i}[(-\infty,\infty)\otimes \{{\bf
x}_i\}],\quad
\quad
\gamma_2=\cup_j (-1)^{\pi_j}[\{t_2\}\otimes S^2_{r_j}],
\end{equation}
$(-1)^{\sigma_i}$ and $(-1)^{\pi_j}$ denoting the orientations of the
various
components of $\gamma_1$ and $\gamma_2$, one gets
\begin{equation}
{\cal A}=128 \pi^4 \sum_{ij}(-1)^{\sigma_i+\pi_j}[1+H(r_j-x_i)],
\label{am}
\end{equation}
i.e.\ ${\cal A}$ counts the link number.

Note that the numerical coefficients in (\ref{conf}) and
(\ref{conf2}) are the
same
(the eventual sign depends on the orientations of the cycles). This
means that
the observables (\ref{obs}) have a correct relative normalization. As
a matter
of fact, due to $256\pi^4=(16\pi^2)^2$, we find that the normalized
expression
generating the full set of observables is
\begin{equation}
{1\over 16\pi^2}\hat F^a\hat F^a.
\end{equation}
This agrees with the normalization of the instanton number:
${1\over 16\pi^2}\int_{M}F^aF^a=1$. Amplitude (\ref{am}) becomes
$1/2\, \sum_{ij}(-1)^{\sigma_i+\pi_j}[1+H(r_j-x_i)]$.
The fact that we have noticed has not to be underestimated: it seems
that the
moduli space is made exactly to adjust the factors, so that,
once the instanton number is normalized correctly, there is no more
freedom in
topological Yang-Mills theory.

One can think of computing links in much more complicated situations,
where, for example,
$\gamma_2$ is a genus $g$ Riemann surface.

\subsection{Conclusion}

The computations made up to now should suffice to show that  ``step
amplitudes'' are part of life when dealing with instanton calculus in
non-abelian Yang-Mills theory.

The richness of the homology of $M\backslash (\gamma_{i_1}\cup\ldots
\gamma_{i_n})$ suggests that there is a
fully open problem, in contrast with the naive expectations of an
empty theory, suggested by the well-known claims. Thus, topological
Yang-Mills
theory is not just Donaldson theory, rather
it also contains a {\sl link} theory. The definition of the map $\pi$
has to be
corrected by
taking into account that the important homology is that of
$M\backslash
(\gamma_{i_1}\cup\ldots \gamma_{i_n})$ (and so it depends on the
amplitude) and
not simply that of $M$.

For higher instanton number $k$, there are many more observables. One
can
construct them from identities like $d\, ({\rm tr}[F\wedge
\cdots\wedge
F])^n=0$. In the $k=1$ case, it is easily proved, by using
(\ref{hal}), that
$(\hat F^a\hat F^a)^n=0$ $\forall n>1$. Instead, for $k=2$ there are
$13$
moduli
and, for example, one has amplitudes like
\begin{equation}
<\phi^a\phi^a(x_1)\cdot \phi^b\phi^b(x_2)\cdot \phi^c\phi^c(x_3)\cdot
{\cal
O}^{(3)}_{\gamma_3}>.
\end{equation}
Here there are three marked points, plus the one at infinity, so
there are
three independent noncontractible 3-spheres $\gamma_3$.
If one inserts two ${\rm tr}[\phi^2]$ operators, one marks two
points. Then
there are two nontrivial three-cycles $S^{(3)}$ and $S^{(3)\prime}$.
Descent
equation from $d (\hat F^a\hat F^a)^2=0$ produce, in particular, a
five
dimensional forms $\omega^{(5)}$ obtained by integrating over one of
these
3-cycles $\gamma_3$:
\begin{equation}
\omega^{(5)}_{\gamma_3}=4\int_{\gamma_3}F^a\psi^a\phi^b\phi^b+2
F^a\phi^a\psi^b\phi^b+\psi^a\psi^a\psi^b\phi^b.
\end{equation}
So, good amplitudes are
\begin{equation}
<\phi^a\phi^a(x_1)\cdot \phi^b\phi^b(x_2)\cdot
\omega^{(5)}_{\gamma_3}>.
\end{equation}
The selection rule can also be saturated with one local observable
${\rm tr}[\phi^2](x)$ and a nonlocal observables
$\omega^{(9)}_{\gamma_3}$
coming from the descent equations generated by
$(\hat F^a \hat F^a)^3$, again integrated over a nontrivial 3-cycle
$\gamma_3$. Lots of other amplitudes, that I do not list here, come
from the
couples $(\gamma_1,\gamma_2)$ of 1-cycles $\gamma_1$ and 2-cycles
$\gamma_2$.

The solution to the problem, that would seem very difficult
at first, could instead be simple, to some extent .
Presumably, one can find a set of recursion relations relating the
amplitudes
with different values of $k$ and some kind of hierarchy collecting
the full set
of them.
In particular, what is the role of $k$ in connection with link
theory?
It would also be interesting to know what happens with
other gauge-groups $G$, in particular $SU(3)$ and to find out
the general characterization of topological Yang-Mills theory.

This concludes the discussion on topological Yang-Mills theory.
In the next section, I turn to topological gravity.

\section{Topological gravity}
\label{tg}
\setcounter{equation}{0}

In this section, I explore topological gravity with the Eguchi-Hanson
instanton, $M=T^*(P_1({\bf C}))$
\cite{eguchi,eguchivarie}.
I parametrize the vierbein with three moduli as follows
\begin{equation}
e^0=\sqrt{\rho^2+a^2\over \rho^2+2 a^2}d\rho,\quad
e^1=\sqrt{\rho^2+a^2}\sigma_x^\prime,\quad
e^2=\sqrt{\rho^2+a^2}\sigma_y^\prime,\quad
e^3=\rho\sqrt{\rho^2+2 a^2\over \rho^2+a^2}\sigma_z^\prime.
\label{coor}
\end{equation}
$a$ is one modulus, namely the size of the instanton. $\rho\in
(0,\infty)$
is related to the usual radial coordinate $r$ by $r^2=\rho^2+a^2$.
It is better to avoid using $r\in (a,\infty)$, because its range
is $a$-dependent and this affects the differentiation with respect to
$a$ and
the parametrizations
of cycles (for example, the 3-sphere $r=$constant would depend on
$a$).

$\sigma_x^\prime$, $\sigma_y^\prime$ and $\sigma_z^\prime$ are
one-forms
satisfying the $SU(2)$ Maurer-Cartan equations
\begin{equation}
d\sigma_i^\prime=\varepsilon_{ijk}\sigma_j^\prime\sigma_k^\prime
\label{cartan}
\end{equation}
and can be expressed as
\begin{eqnarray}
\sigma_x^\prime&=&\cos \beta \, \sigma_z-\sin \alpha\, \sin \beta
\,\sigma_x
-\cos \alpha \, \sin \beta\,\sigma_y={1\over 2}(\sin\psi^\prime\,
d\theta^\prime-\sin \theta^\prime \,\cos \psi^\prime  \, d\phi^\prime
),\nonumber\\
\sigma_y^\prime&=&\cos \alpha \, \sigma_x-\sin \alpha
\,\sigma_y=-{1\over
2}(\cos \psi^\prime \, d\theta^\prime +\sin \theta^\prime  \, \sin
\psi^\prime
\, d\phi^\prime ),\nonumber\\
\sigma_z^\prime&=&\sin \beta\, \sigma_z+\sin \alpha \cos
\beta \,\sigma_x +\cos \alpha \,\cos \beta \,\sigma_y=
{1\over 2}(d\psi^\prime +\cos \theta^\prime \, d\phi^\prime ).
\end{eqnarray}
$\sigma_i^\prime$ is nothing but a rotated version of the usual basis
$\sigma_i$ \cite{eguchi}:
\begin{eqnarray}
\sigma_x&=&{1\over 2}(\sin\psi\, d\theta-\sin \theta \,\cos \psi \,
d\phi),\nonumber\\
\sigma_y&=&-{1\over 2}(\cos \psi\, d\theta+\sin \theta \, \sin \psi
\,
d\phi),\nonumber\\
\sigma_z&=&{1\over 2}(d\psi+\cos \theta\, d\phi).
\end{eqnarray}
One can write $\theta^\prime$, $\psi^\prime$ and $\phi^\prime$
as functions of  $\theta$, $\psi$, $\phi$ and the moduli $\alpha$ and
$\beta$. Nevertheless,
the ranges are the same for unprimed as for primed angles:
\begin{equation}
0\leq \theta\leq \pi,\quad\quad 0\leq \phi\leq 2\pi,\quad\quad 0\leq
\psi\leq
2\pi.
\label{rangles}
\end{equation}
The angles $\alpha$ and $\beta$ have been introduced to deal
explicitly with a three dimensional moduli space.
The metric is self-dual for any $\alpha$ and $\beta$, because
in  proving self-duality, one only needs to use (\ref{cartan}).
The relations (\ref{cartan}) are preserved by rotations in
the space of the forms
$\sigma_i$. However, $\sigma_x$ and
$\sigma_y$ appear symmetrically in the metric $ds^2=e^ae^a$,
so that one of the three rotations (the one around the $z$ axis)
is a Lorentz rotation and
not a true modulus. So, one remains precisely with two meaningful
angles. This matches with
the counting of the number of moduli in the Gibbons-Hawking
\cite{gibbons} description of multicenter metrics.

To some extent, the number of moduli can be a convention. One can say
that $\alpha$ and $\beta$ are trivial parameters and not moduli,
because they can be eliminated with a change of coordinates.
However, such a change of coordinates is a rotation and so is
not bounded at infinity.
A ghost field tending to a constant at infinity (i.e.\
to a global gauge transformation) should be allowed.
Nevertheless, from the examples of the previous and the present
section,
it seems reasonable to restrict the
ghost fields to be bounded at infinity.

Moreover, the theory of topological gravity that we are considering,
is related {\sl via}
topological twist to N=2 supergravity \cite{mepie} and it seems
that in supergravity the full set of moduli should be treated
\cite{konishi}.
So, in order
to develop a background for a
future comparison with N=2 supergravity computations \cite{romani},
I keep the full set of three moduli.

The moduli space we are dealing with is ${\cal M}=({\bf
R}^3\backslash\{0\})/{\bf Z}_2$.
This can be easily seen in the Gibbons-Hawking description
\cite{gibbons},
where the metric is identified by the positions of two non-coincident
points
in ${\bf R}^3$: one point can always be placed in the origin (with a
translation)
and the quotient by ${\bf Z}_2$ means that {\sl which} one is placed
there
is immaterial. ${\cal M}$ can also be written as
$(0,\infty)\otimes S^2/{\bf Z}_2$,
the radial coordinate being $a\in (0,\infty)$ and the polar angles of
$S^2$
being $\alpha$ and $\beta$, which have ranges
\begin{equation}
0\leq \alpha\leq 2\pi,\quad\quad -{\pi\over 2}\leq \beta\leq
{\pi\over 2}.
\end{equation}
Indeed, for $\beta=\pm {\pi\over 2}$ the metric is
$\alpha$-independent, so
that
 $\beta=\pm {\pi\over 2}$ are the two poles of the sphere $S^2$. The
quotient
by ${\bf Z}_2$
will be taken into account by dividing the results by a factor 2.

For future use, let me write down the spin connection and the
curvature
explicitly
\begin{equation}
\matrix{
\omega^{01}=-\omega^{23}=-\rho{\sqrt{\rho^2+2a^2}\over
\rho^2+a^2}\sigma_x^\prime,
&R^{01}=-R^{23}={2a^4\over (\rho^2+a^2)^3}(-e^0e^1+e^2e^3),\cr
\omega^{02}=-\omega^{31}=-\rho{\sqrt{\rho^2+2a^2}\over
\rho^2+a^2}\sigma_y^\prime,
&R^{02}=-R^{31}={2a^4\over (\rho^2+a^2)^3}(-e^0e^2+e^3e^1),\cr
\omega^{03}=-\omega^{12}=-\left[1+{a^4\over
(\rho^2+a^2)^2}\right]\sigma_z^\prime,
&R^{03}=-R^{12}=-{4a^4\over (\rho^2+a^2)^3}(-e^0e^3+e^1e^2).}
\end{equation}

Now, we are ready to compute observables and amplitudes.

\subsection{The simplest amplitude}

We begin with the calculation of the easiest observable,
namely ${\cal O}^{(3)}_{\gamma_3}$ of (\ref{grobs}).
The simplest choice of $\gamma_3$ is a 3-sphere
$S^3_\rho$ of radius $\rho$, even if naively it is not the
representative of a
3-cycle and so it is expected to produce a trivial result, i.e.\ an
exact form
$\omega^{(1)}_\rho={\cal O}^{(3)}_{S^3_\rho}$.
One can write
\begin{equation}
{\cal O}^{(3)}_{S^3_\rho}=2\int_{S^3_\rho}R^{ab}\chi^{ab}=
2\int_{S^3_\rho}R^{ab}\chi^{\prime ab}.
\end{equation}
The last equality follows from the fact that
$R^{ab}(\chi^{ab}-\chi^{\prime
ab})=R^{ab}{\cal D}\varepsilon^{ab}$ is a total derivative,
due to the Bianchi identity ${\cal D}R^{ab}=0$.
Now, $\chi^{\prime ab}$ is easily computed, since (\ref{brsgrav})
shows that it
is the exterior derivative of
$\omega^{ab}$ with respect to the moduli:
\begin{eqnarray}
\chi^{\prime 01}&=&-\chi^{\prime 23}={2\rho a^3 da\,
\sigma_x^\prime\over
(\rho^2+a^2)^2\sqrt{\rho^2+2a^2}}+
\rho{\sqrt{\rho^2+2a^2}\over \rho^2+a^2}(d\beta\,
\sigma_z^\prime+\sin \beta\,
d\alpha\, \sigma_y^\prime),\nonumber\\
\chi^{\prime 02}&=&-\chi^{\prime 31}={2\rho a^3 da\,
\sigma_y^\prime\over
(\rho^2+a^2)^2\sqrt{\rho^2+2a^2}}+
\rho{\sqrt{\rho^2+2a^2}\over \rho^2+a^2}d\alpha\,
(\cos\beta\,\sigma_z^\prime-\sin \beta\, \sigma_x^\prime),\nonumber\\
\chi^{\prime 03}&=&-\chi^{\prime 12}=-{4\rho^2 a^3 da\,
\sigma_z^\prime\over
(\rho^2+a^2)^3}-\left[1+{a^4\over (\rho^2+a^2)^2}\right]
(d\beta\,\sigma_x^\prime+\cos \beta\, d\alpha\, \sigma_y^\prime),
\end{eqnarray}
The straightforward computation leads to
\begin{equation}
\omega^{(1)}_{\rho}=-192{a^7\rho^2da\over (\rho^2+a^2)^5}
\int_{S^3_\rho}\sigma_x^\prime\sigma_y^\prime\sigma^\prime_z=192\pi^2{
a^7\rho^2da\over (\rho^2+a^2)^5}.
\label{forma}
\end{equation}
We can also write
\begin{equation}
\omega^{(1)}_{\rho}=df,
\quad\quad
f(\rho,a)=-24\pi^2\rho^2{4a^6+6a^4\rho^2+4a^2\rho^4+\rho^6\over(a^2+\rho^2)^4}.
\end{equation}
In $df$, the exterior derivative $d$ acts on $a$. We see that
$\omega^{(1)}_\rho$
is a well behaving 1-form on the moduli space ${\cal M}$.
Integrating it over the 1-cycle $\gamma_1\subset {\cal M}$,
$\gamma_1=(0,\infty)\times\{\hat n\}$, $\hat n$ denoting a certain
point of
$S^2/{\bf Z}_2$, one gets a finite $\rho$- and $\hat n$-independent
result:
\begin{equation}
\int_{\gamma_1}\omega^{(1)}_{r}=24\pi^2,
\label{res}
\end{equation}
which can also be considered as the amplitude $<{\cal
O}^{(3)}_{S^3_\rho}>$,
when one restricts the moduli space to be only the range $(0,\infty)$
of the
scale $a$.

The above result means once again that an amplitude naively
expected to be zero is instead nonvanishing.
So, it is an anomaly. Nevertheless, we should find a more
satisfactory
explanation, since there is no other observable around that can mark
a cycle
and change the topology of the Eguchi-Hanson manifold.
Indeed, (\ref{res}) is always different from zero: there is no step
function.
So, what is the reason why the 3-sphere $S^3_\rho$ should be
considered indeed
a 3-cycle?
Once again, the whole story is originated by the boundary of the
moduli space.
Consider the limiting case $a=0$. There, the Eguchi-Hanson manifold
degenerates
to ${\bf C}^2/{\bf Z}_2$, which is singular in the origin. The
consequence is
that if one wants to
shrink $S^3_\rho$, one has necessarily to cross the singular point.
Nothing can be said a priori about this procedure: it could be safe,
or produce
infinities, or, in the
most interesting case, produce finite nonvanishing anomalous values,
as it happens in fact.

As it was partially anticipated in the previous section, one can say
that the
above anomalies
are not, strictly speaking, ``anomalies''.
Indeed, if an object is not BRST exact on the entire moduli space
(boundary included), then it should not be considered as BRST exact.
So, one
should simply
change the naive definition of BRST cohomology accordingly.

The check of  the independence of result (\ref{res}) from the
coordinate system
is left to the appendix. There, I use the Gibbons-Hawking
coordinates. This
check is interesting for preparing the future computations in
multi-center
metrics, because it reveals some technical subtleties that arise
due to the presence of Dirac strings.

The reason for the above result to be necessarily nonvanishing is
quite simple.
Indeed, the observable ${\cal O}^{(3)}$ can be written, like in
formula
(\ref{417}), in terms of the Chern-Simons form. In other words, the
function
$f(\rho,a)$
is precisely the integral of the Chern-Simons form over
$S^{(3)}_\rho$. Since
in our example
only two dimensionful parameters are around ($a$ and $\rho$), one can
write
$f(\rho,a)=f(\rho/a)$ and
conclude
\begin{equation}
<{\cal O}^{(3)}_{S^3_\rho}>=\int_0^\infty{\partial f\over \partial
a}da=
f(\rho/a)\Big|^{a=\infty}_{a=0}=-
f(\rho/a)\Big|^{\rho=\infty}_{\rho=0}=-\int_0^\infty {\partial f\over
\partial
\rho}d\rho.
\end{equation}
The last expression is proportional to the topological invariant
$\int_M {\rm
tr}[R\wedge R]$
on the Eguchi-Hanson manifold $M$. So, it is surely nonzero.
This is an aspect of the fact that
many topological invariants of moduli space ${\cal M}$ computed by
the
topological field theory
are related to topological invariants of the manifold $M$ (see for
example
\cite{wittenN=1}).
One can turn this argument backwards and say: since the Eguchi-Hanson
manifold
has no 3-cycles (for generic $a$),
but the above amplitude {\sl must} be nonzero
(because the Pontrjiagin number is nonvanishing),
then {\sl necessarily} there must be some singularity
in the boundary of the moduli-space. In other words,
it would have been impossible for the Eguchi-Hanson instanton
to have a regular limit for $a\rightarrow 0$. This seems a general
property of instantons and, perhaps, some similar argument
could give general information about the
singularities that the moduli space of instantons {\sl necessarily}
possesses.

When considering the full three dimensional moduli space, one has to
deal with
a more complicated amplitude, as we are going to see.

\subsection{Solution}

In order to compute the remaining observables, we have to work out
the explicit
expressions of some ghosts. We have
\begin{equation}
\psi^{\prime ab}=\left(\matrix{
-{a\rho^2\,da\over (\rho^2+a^2)(\rho^2+2 a^2)}&0&0&0\cr
0&{a\,da\over \rho^2+a^2}&-\sin \beta \, d\alpha &-K d\beta\cr
0&\sin \beta\, d\alpha&{a\,da\over \rho^2+a^2}&-K\cos\beta\,
d\alpha\cr
0&d\beta/K&\cos\beta\, d\alpha/K&{a\rho^2\,da\over
(\rho^2+a^2)(\rho^2+2
a^2)}}\right)=
\psi^{ab}+{\cal D}^b\varepsilon^a-\varepsilon^{ab},
\end{equation}
where $K={\rho^2+a^2\over \rho\sqrt{\rho^2+2 a^2}}$. Note the bad
divergence at
$\rho=0$,
that, however, is not dangerous, since
it will disappear at the end.  Indeed, the ghost
$\varepsilon^a$, although being regular, subtracts the divergence in
question
from $\psi^{\prime ab}$,
so that $\psi^{ab}$ turns out to be also regular.
It is worth to stress again that this is due to the correct choice
of the gauge-fixing conditions that break the gauge of the gauge.

The equation for $\varepsilon^a$ will be solved in two steps. In
general,
$\varepsilon^a$
is a linear combination  of $da$, $d\alpha$ and $d\beta$. The
coefficient of
$da$ and those of $d\alpha$, $d\beta$ satisfy independent equations
and they
will be discussed independently. So, in the first step, we put
$d\alpha=d\beta=0$ and choose the following ansatz
\begin{equation}
\varepsilon^0=f(\rho,a){\rho\over a}\sqrt{\rho^2+a^2\over \rho^2+2
a^2}da,\quad\quad \varepsilon^1=\varepsilon^2=\varepsilon^3=0.
\end{equation}
The calculation goes on straightforwardly. One has to combine the
first of
(\ref{gfix}) with the first
of (\ref{gfsol}). This produces the following equation for $F=1+f$:
\begin{equation}
F^{\prime\prime}+{5\rho^2+3a^2\over
\rho(\rho^2+a^2)}F^\prime+{4a^2\over
(\rho^2+a^2)^2}F=0.
\end{equation}
The ansatz
\begin{equation}
F=\rho^m(\rho^2+a^2)^n,
\end{equation}
gives $m=-2$, $n=\pm 1$. The coefficients of the linear combination
of the
corresponding two solutions have to be chosen in order to have a
regular
$\varepsilon^0$ everywhere and with a regular limit at
infinity. This produces the answer
\begin{equation}
f(\rho,a)={a^2\over \rho^2+a^2}.
\end{equation}
Note the $1\over \rho$ behavior of $\varepsilon^0$ at
$\rho\rightarrow \infty$
that we have already encountered when discussing $C^a$.

In the second step of the calculation, we put $da=0$ and choose the
ansatz
\begin{equation}
\varepsilon^0=0,\quad \varepsilon^1=-f_1\sqrt{\rho^2+a^2}\cos \beta\,
d\alpha,\quad
\varepsilon^2=f_2 \sqrt{\rho^2+a^2}d\beta,\quad \varepsilon^3=0.
\end{equation}
It is then easy to see that $f_1$ and $f_2$ satisfy independent
equivalent
equations, so that the choice $f_1=f_2=f$ is consistent. One finds
the equation
\begin{equation}
(f^\prime L)^\prime={2 a^8\over L}(1+2f),
\end{equation}
where $L=\rho(\rho^2+a^2)(\rho^2+2a^2)$. Setting $F=1+2f$,
$G=F^\prime L$, one
has
\begin{equation}
GG^\prime=4a^8FF^\prime,
\end{equation}
which is solved by $F={1\over 2a^4}G$. Integrating this last equation
and
choosing the constant so as to have a bounded $\varepsilon^a$ for
$\rho\rightarrow\infty$, one finally finds
\begin{equation}
f={1\over 2}\left(
\rho{\sqrt{\rho^2+2a^2}\over \rho^2+a^2}-1\right).
\end{equation}
Collecting the whole information that we have accumulated so far, the
final
result is
\begin{eqnarray}
\varepsilon^0&=&{\rho a da\over
\sqrt{(\rho^2+a^2)(\rho^2+2a^2)}},\quad\quad
\varepsilon^1=-{1\over 2}\cos\beta\, d\alpha\left(\rho
\sqrt{\rho^2+2 a^2\over
\rho^2+a^2}-\sqrt{\rho^2+a^2}\right),\nonumber\\
\varepsilon^2&=&{1\over 2}d\beta\left(\rho
\sqrt{\rho^2+2 a^2\over
\rho^2+a^2}-\sqrt{\rho^2+a^2}\right),\quad\quad
\varepsilon^3=0.
\end{eqnarray}

We have thus arrived at the explicit expression of $\varepsilon^{ab}$
which is
all what we need:
\begin{eqnarray}
\varepsilon^{01}&=&-{1\over 2}\cos\beta\, d\alpha
\left(\rho{\sqrt{\rho^2+2
a^2}\over \rho^2+a^2}-1\right),
\quad \varepsilon^{02}={1\over 2}d\beta \left(\rho{\sqrt{\rho^2+2
a^2}\over
\rho^2+a^2}-1\right),\quad
\varepsilon^{03}=0,\nonumber\\
\varepsilon^{12}&=&\sin \beta\, d\alpha,\quad
\varepsilon^{13}={1\over 2}d\beta
\left(\rho{\sqrt{\rho^2+2 a^2}\over \rho^2+a^2}+1\right),\quad
\varepsilon^{23}=
{1\over 2}\cos \beta\, d\alpha\left(\rho{\sqrt{\rho^2+2 a^2}\over
\rho^2+a^2}+1\right).\nonumber\\
\label{varepsilonab}
\end{eqnarray}
Note that there is no $da$ in the expression of $\varepsilon^{ab}$. I
do not
write down the explicit
expression of $\psi^{ab}$, nevertheless it can be easily checked that
it is not
singular at $\rho\rightarrow 0$, as promised.

\subsection{Other amplitudes}

We are now ready to discuss observables and amplitudes.
Let us start from the observable ${\cal O}^{(2)}_{\gamma_2}$. It is
well-known
that
multicenter manifolds possess nontrivial 2-cycles \cite{yuille}.
In particular, the Eguchi-Hanson manifold
possesses one noncontractible 2-sphere $S^2$. In the usual
Gibbons-Hawking
description
\cite{gibbons}, multicenter metrics are parametrized by the positions
of $n$
points (centers)
in ${\bf R}^3$ and by a cyclic coordinate $\tau\in [0,4 \pi]$. The
nontrivial
2-spheres $S^2$ are represented by lines $l$ in ${\bf R}^3$ joining
couples of
centers,
``multiplied'' by the full range of $\tau$. The line $l$ can be
thought as a
meridian on $S^2$, while the cyclic coordinate $\tau$ is the
longitudinal angle. There are also noncompact 2-cycles: for example,
one can
choose a line $l$ that goes from one of the centers to infinity.

Before going on, we have to pay attention to the following fact:
the line $l$ that joins the two centers of the Eguchi-Hanson manifold
depends
on the positions on the centers, so it seems difficult to parametrize
$\gamma_2$ in a moduli-independent way. If a cycle $\gamma$ is
parametrized in
a moduli-dependent way, one cannot interchange the BRST operator with
the
integral over the cycle and prove that (\ref{obs}) or (\ref{grobs})
are BRST
closed.
In the Gibbons-Hawking coordinates, there is at least one noncompact
2-cycle that can be parametrized in a moduli-independent way:
it is described by a line $l$ going from the center placed in the
origin to infinity. Instead, in our coordinates it is easy to see
that
the two centers of the Eguchi-Hanson metric correspond to $\rho=0$,
$\theta^\prime=0,\pi$.
So, we can parametrize a class of representatives of the 2-cycle
$\gamma_2$ by
\begin{equation}
S^2_{g}:\quad\quad\rho=g(\theta^\prime ),\quad \theta^\prime
\in[0,\pi],\quad
\phi^\prime \in[0,2\pi],\quad
\psi^\prime ={\rm const},
\label{class}
\end{equation}
$g(\theta^\prime )$ being an arbitrary function such that
$g(0)=g(\pi)=0$.
This is a moduli-dependent parametrization, because the primed angles
depend on
$\alpha$ and $\beta$.
So, we are not guaranteed to compute meaningful quantities.
Eventually, one should change variables from $\theta,\phi,\psi$ to
$\theta^\prime,
\phi^\prime,\psi^\prime$, according to the rules explained
at the end of section \ref{formalism}\footnotemark\footnotetext{This
change of
variables makes
the new metric $\alpha,\beta$-independent.
Then, $d\alpha$ and $d\beta$ are the coefficients of two unbounded
zero
modes of $\varepsilon^a$. This is due to the peculiarity of the
two moduli $\alpha$ and $\beta$.}.

Instead of doing this lengthy work, we can disentangle the situation
as follows
(it should be kept in mind the what follows is an {\sl ad hoc}
argument,
differently from the other arguments contained in the present paper).

Our purpose is to compute the amplitude
\begin{equation}
{\cal A}=<{\cal O}^{(3)}_{S^3_\rho}\cdot {\cal
O}^{(2)}_{S^2_g}>=\int_{\cal
M}\omega^{(1)}_\rho\wedge \omega^{(2)}_g,
\label{ampli}
\end{equation}
$\omega^{(1)}_\rho$ being given by (\ref{forma}) and $\omega^{(2)}_g$
being
${\cal O}^{(2)}_{S^2_g}$. Since $\omega^{(1)}_\rho$ is proportional
to $da$, we
can work out
$\omega^{(2)}_g$ at fixed $a$. The amplitude can be rewritten as
\begin{equation}
{\cal A}=\int_{\cal M}df\wedge \omega_g^{(2)}=\int_{\partial {\cal
M}}f\omega^{(2)}_g={1\over 2}24\pi^2\lim_{a\rightarrow
0}\int_{S^2}\omega_g^{(2)}.
\end{equation}
The factor $1/2$ is due to the quotient by ${\bf Z}_2$.

We shall use the parametrization (\ref{class}), although it is
moduli-dependent. At fixed $a$,
$\omega^{(2)}_g$ is surely closed, because it is a top form of
$S^2\in{\cal
M}$. Moreover,
${\cal A}$ should not depend on the representative of $\gamma_2$,
since,
although it is necessary to use the moduli for parametrizing the line
$l$
joining the two centers, this is no longer true for the {\sl
difference}
$\delta\gamma_2$ of two representatives of $\gamma_2$: indeed,
$\delta\gamma_2$
is represented by a loop $\delta l$  in ${\bf R}^3$. This permits to
repeat
safely the proof that adding a boundary to $\gamma_2$ amounts to
adding an
exact form to $\omega^{(2)}$. Summarizing, we have reasons to expect
that the
above amplitude is well-defined and topological, notwithstanding the
unconventional parametrization of $\gamma_2$.

We have to be careful before taking the limit $a\rightarrow 0$, since
${\cal
O}^{(2)}_{S^2_g}$ contains expressions like
\begin{equation}
I=\int_{0}^\pi{a^8 g(\theta^\prime) dg(\theta^\prime) \, \cos
\theta^\prime\over (g(\theta^\prime)^2+a^2)^5}
\end{equation}
(the cyclic coordinate $\phi$ having already been integrated away),
coming from terms containing $e^0$. For $a\rightarrow 0$, one finds
$0\times\int {dg\over g^9}$, which is surely problematic, since
$g(0)=g(\pi)=0$. We can overcome the difficulty with an integration
by parts:
\begin{equation}
I=-{1\over 8}\left[{a^8 \cos\theta^\prime\over
(g(\theta^\prime)^2+a^2)^8}\right]^\pi_0+{1\over 8}\int_{0}^\pi{a^8
\sin
\theta^\prime\, d\theta^\prime \over
(g(\theta^\prime)^2+a^2)^4}={1\over
4}+{1\over 8}\int_{0}^\pi{a^8 \sin \theta^\prime\, d\theta^\prime
\over
(g(\theta^\prime)^2+a^2)^4}.
\label{contr}
\end{equation}
The second term tends to zero, thus the net result is a finite
contribution. As
a matter of fact, one can show that the only nonzero contributions to
${\cal
A}$ are originated in this way.
Collecting all the terms together, one ends up with
\begin{equation}
{\cal A}=24\pi^2\cdot16\pi^2.
\label{am1}
\end{equation}
We have thus made a very powerful test of topological field theory:
we have
found a topological amplitude that is not only independent of some
parameters,
but of a
full function $g(\theta^\prime)$.

The above correlation function has a simple geometrical
interpretation.  Apart
from the quotient with respect to ${\bf Z}_2$, the moduli space is
${\bf
R}^3\backslash \{0\}$. $\omega^{(1)}$ is the Poincar\`e
dual of a noncontractible 2-sphere $H_2$ around the origin, while
$\omega^{(2)}$ can be thought as the Poincar\`e dual of  a noncompact
1-cycle
$H_1$ joining the point $0$
with infinity. It is clear that the intersection of these two cycles
on ${\cal
M}$ is a point:
\begin{equation}
\#(H_1\cap H_2)=1.
\end{equation}
Our amplitude can be interpreted as this (the normalization is
immaterial, at
this level). In other words,
thanks to the anomaly that we have discovered, the topological field
theory is
able to
reproduce some intersection theory on the moduli space.

Note the following funny fact: $\omega^{(2)}$ was calculated on a
compact cycle
of the manifold $M$,
but it gives a noncompact cycle on the moduli space ${\cal M}$.
Instead,
$\omega^{(1)}$
relates a compact cycle of $M$ with a compact cycle of ${\cal M}$.
Thus, the map $\pi$ mixes compact and non-compact homologies in a
nontrivial
way.

One can also consider noncompact  2-cycles like
\begin{equation}
\gamma^{(2)}_{g}:\quad\quad\rho=g(\theta^\prime ),\quad \theta^\prime
\in[0,\theta^\prime_{max}],\quad \phi^\prime \in[0,2\pi],\quad
\psi^\prime ={\rm const},
\label{class2}
\end{equation}
with $g(0)=0$, $g(\theta^\prime_{max})=\infty$. Then, since the full
result is
encoded in expressions like (\ref{contr}), one finds a half than
before:
\begin{equation}
{\cal A}=24 \pi^2\cdot 8\pi^2.
\label{am2}
\end{equation}

The correct normalization seems to be
obtained by dividing by $24 \pi^2\cdot 8\pi^2$, so that the
two amplitudes (\ref{am1}) and (\ref{am2}) become $2$ and $1$,
respectively.
Such numbers are not so surprising, since (modulo signs) they are the
entries
of the Cartan matrix of the simply laced groups
\cite{ale}. The phenomenon that we noticed in topological Yang-Mills
theory,
i.e.\ the strict relation among the normalizations of the observables
and the
meaningful topological invariant, does not seem to occur in
topological
gravity. Notice, however, that in topological gravity the observables
are
related to the Pontrjiagin number, while the meaningful
topological invariant is the Hierzebruch signature, that differs from
the
Pontrjiagin number by boundary corrections.

Actually, there could be another nontrivial amplitude, since there is
another
nonvanishing intersection number: this is the intersection between
the full
manifold ${\cal M}$ (3-cycle)
and a point $x$ (zero-cycle),
\begin{equation}
\#(\{x \}\cap {\cal M})=1.
\label{num1}
\end{equation}
The observable that is Poincar\`e dual to ${\cal M}$ is the
identity operator, while the observable that is Poincar\'e dual to a
point
should be a 3-form
on ${\cal M}$. Thus, the natural candidate for the latter is ${\cal
O}^{(1)}_{\gamma_1}$. One natural 1-cycle $\gamma_1$ (a circle)
that comes to one's mind is the following: in the Gibbons-Hawking
description,
fix a point ${\bf x}$  in ${\bf R}^3$ and take the parameter of
$\gamma_1$
to be the cyclic coordinate $\tau\in[0,4\pi]$. Such a circle will be
denoted by
$\gamma_{\bf x}$. This is surely a contractible cycle, as far as $a$
in
nonzero. Indeed,
there exists a line $l$ joining the two centers and passing through
${\bf x}$.
Such line, ``multiplied'' by the range of $\tau$, represents a
non-contractible
2-sphere $S^2_l$, as already recalled.
$\tau$ is the longitudinal angle on it. Then, one can deform
$\gamma_{\bf x}$
on $S^2_l$ and make it collapse on the northern or the southern poles
(the
centers). This corresponds to
move ${\bf x}$ to one of the two centers.
However, when $a$ is $0$ the two centers coincide, i.e.\
northern and southern poles of $S^2_l$ coincide $\forall l$.
What we learned in the computations that we have done so far tell us
that we
should be cautious before drawing conclusions. We simply trust in the
physical
formalism,
that should already contain the correct answer.

Since we are not using the Gibbons-Hawing coordinates, let us choose
a
convenient 1-cycle $\gamma_1$ in our parametrization. Now, $\tau$
corresponds
to $\phi/2$ (indeed, $\phi$ is the cyclic coordinate of out metric).
So, let us take
\begin{equation}
\gamma_1:\quad\quad \rho=\rho_0,\quad \theta=\theta_0, \quad
\psi=\psi_0,\quad
\phi\in[0,2\pi].
\end{equation}
One then finds that the $\alpha$ and $\beta$-integrations kill every
term, and
the result is zero.
Actually, it must be so, since $d\phi$ always appears multiplied by
$\cos\theta$ or by $\sin\theta$:
a nonzero result would be $\theta$-dependent. As a nontrivial check
of this and
of the consistency of the computational methods that we have used so
far, let
us take a moduli-dependent parametrization
\begin{equation}
\gamma_1:\quad\quad \rho=\rho_0,\quad \theta^\prime=\theta^\prime_0,
\quad
\psi^\prime=\psi^\prime_0,\quad \phi^\prime\in[0,2\pi],
\end{equation}
and write the observable as
\begin{equation}
{\cal
O}^{(1)}_{\gamma_1}=s\int_{\gamma_1}2\varepsilon^{ab}\chi^{\prime
ab}+\varepsilon^{ab}{\cal D}\varepsilon^{ab},
\end{equation}
(notice that the BRST operator $s$ is {\sl outside} the integral, to
avoid
problems with the moduli-dependent parametrization).
Now,  the $\alpha$ and $\beta$-integrations do not kill anything,
and it is a nontrivial cancellation that makes the total to be zero.
So, apparently the theory is not able to reproduce the intersection
number
(\ref{num1}). I do not have any better explanation of this fact.
Something
more on this will be said in the next section.

Summarizing, in topological gravity, we have found consistent
amplitudes. Another question remains without answer:
does topological gravity contain a kind of link theory?
Can we find step amplitudes?

\section{Coupled theories}
\label{couple}

In this section, I consider topological gravity on the Eguchi-Hanson
manifold
coupled to abelian topological Yang-Mills theory. Indeed, there is
exactly one
self-dual $U(1)$ gauge-connection on the Eguchi-Hanson manifold,
namely
\begin{equation}
A={2a^2\over \rho^2+a^2}\sigma_z^\prime,
\end{equation}
the normalization being such that $\int_M F\wedge F=8\pi^2$.
Moreover, $A=A_ae^a$ is gauge-fixed with the condition ${\cal
D}_aA^a=0$.
Applying the method of section \ref{formalism}, one finds
\begin{equation}
\psi={4a\rho^2da\over (\rho^2+a^2)^2}\sigma_z^\prime+
{2a^2\over \rho^2+a^2}(d\beta\,\sigma_x^\prime+\cos\beta\, d\alpha\,
\sigma_y^\prime),\quad C={\rm const},\quad\phi={\rm const},
\end{equation}
$\psi=\psi_a e^a$ satisfying the condition ${\cal D}_a\psi^a=0$.
The $C$ and $\phi$ zero modes are eliminated by inserting the
puncture operator
\begin{equation}
C(x)\delta[\phi(x)],
\end{equation}
$x$ being any point of $M$. Thus, $\hat F=F+\psi+\phi=F+\psi$.
$F$ can be thought as the Poincar\`e dual of the compact
noncontractible
2-sphere of the Eguchi-Hanson manifold.
In the abelian case, there are observables generated by
$\hat d \, [\hat F^n]=0$.
We can thus compute the amplitude
\begin{equation}
{\cal A}=<\int_{S^3_\rho}\psi^3>=\int_{\cal M}
96\pi^2{a^5\rho^2da\over (\rho^2+a^2)^4}\cos\beta\,d\alpha \, d\beta=
32\pi^3.
\label{la1}
\end{equation}
There is another amplitude, indeed, showing that the coupling between
topological gravity and topological Yang-Mills theory  is nontrivial
{}.

It corresponds to the form
\begin{equation}
\omega^{(3)}=\int_{S^3_\rho}2\eta^{ab}\chi^{ab}
F+\chi^{ab}\chi^{ab}\psi+2
R^{ab}\eta^{ab}\psi,
\end{equation}
generated by $\hat d[\hat R^{ab}\hat R^{ab}\hat F]=0$.
The calculation is a bit long.
The effort can be a bit reduced by writing
$\hat R^{ab}\hat R^{ab}\hat F$ as $\hat d[\hat R^{ab} \hat R^{ab}\hat
A]$ and
\begin{equation}
\omega^{(3)}=d\Omega^{(2)},\quad\quad \Omega^{(2)}=-\int_{S^3_\rho}
(\chi^{ab}\chi^{ab}+2 R^{ab}\eta^{ab})\, A.
\end{equation}
One has to work out $\eta^{ab}$ and $\chi^{ab}$ from
(\ref{varepsilonab})
according to (\ref{brsgrav}).
At the end one finds
\begin{equation}
\Omega^{(2)}=32\pi^2{a^{10}\over (\rho^2+a^2)^5}
\cos\beta\, d\alpha \, d\beta,\quad\quad
{\cal A}=\int_{\cal M} \omega^{(3)}=64\pi^3.
\label{la2}
\end{equation}

One could introduce a parameter $\xi$ in front of $A$, so that
\begin{equation}
A=\xi {2a^2\over \rho^2+a^2}\sigma_z^\prime,\quad\quad \int_MF\wedge
F=8\pi^2\xi^2.
\end{equation}
However, $\xi$ cannot be considered as a modulus,
since it changes the Chern class.
In other words, $s\xi=0$, otherwise the
Chern class in not a good observable.
The integration over $\xi$ is made convergent by the Chern class
itself in the
action:
\begin{equation}
\int d\xi\,{\rm e}^{-8\pi^2\xi^2}.
\end{equation}
Such an integral replaces, in the abelian case,
the usual sum over topological numbers. It gives an overall constant
factor,
that we can normalize to one. This justifies the previous
calculations, in
which $\xi$ was 1.

The amplitudes (\ref{la1}) and (\ref{la2}) could be perhaps related
to the
intersection form
(\ref{num1}), that we were unable to reproduce in pure topological
gravity.

It would also be interesting to work out the relation between the
calculations made in the previous and in the present sections
and the theory obtained by twisting N=2 supergravity \cite{mepie},
in which there is a $U(1)$ gauge connection (graviphoton),
that, however, is related to the ghosts for the ghosts.

\vskip .3in
\begin{center}
{\bf Acknowledgements}
\end{center}

I would like to thank M.\ Bianchi, F.\ Fucito and G.C.\ Rossi for
interesting
discussions
on instanton calculus in supersymmetric theories. This research was
supported
in part by the
Packard Foundation and by NSF grant PHY-92-18167.

\vskip .3in

\section{Appendix: Checks}
\label{app}

In the Gibbons-Hawking multicenter coordinates \cite{gibbons},
one describes the multicenter manifold $M$ as
an ${\bf R}^3$ space,  where the centers are placed, plus a cyclic
coordinate
$\tau$ ranging from $0$ to $4\pi$. One center can always be placed in
the
origin, while the positions of the other centers are the moduli.
The general form of the metric is
\begin{equation}
ds^2=U^{-1}(d\tau+{\bf \omega})^2+U d{\bf x}\cdot d{\bf x}.
\end{equation}
In the Eguchi-Hanson case (two centers),
placing one center in the origin and the other one on the $z$-axis
and choosing
natural vertical positions for the Dirac strings, one has
\begin{equation}
U={1\over |{\bf x}|}+{1\over |{\bf x-a}|}, \quad\quad {\bf
\omega}=\left(
{z\over |{\bf x}|}+{z-a\over |{\bf x}-{\bf a}|}-2\right){ydx-xdy\over
x^2+y^2}.
\end{equation}
We write the observable ${\cal O}^{(3)}_{\gamma_3}$ as
\begin{equation}
{\cal O}^{(3)}_{\gamma_3}=-s\int_{\gamma_3}C,
\end{equation}
$C$ being the Chern-Simons form, that one can easily calculate in
general
(i.e.\ for any multicenter metric), finding the expression
\begin{equation}
C=U^{-9/2}\partial_jU\left(U\partial_i\partial_jU-{3\over
2}\partial_iU\partial_jU\right)\varepsilon_{imn} e^m e^n e^0.
\end{equation}
We choose the cycle $\gamma_3$ as the full range $[0,4\pi]$ of the
cyclic
coordinate $\tau$,
multiplied by a 2-sphere $S^2_r$ of radius $r$
contained in the space ${\bf R}^3$. In particular,
\begin{equation}
\int_{\gamma_3}C=4\pi\int_{S^2_r}U^{-4}\partial_jU\left(U\partial_i\partial_jU-{3\over
2}\partial_iU\partial_jU\right)\varepsilon_{imn}dx^mdx^n\equiv
f(a,r),
\label{inte}
\end{equation}
and our amplitude is
\begin{equation}
{\cal A}=<{\cal O}^{(3)}_{\gamma_3}>=-\int_0^\infty df(a,r).
\end{equation}
The above integration looks trivial, at first sight, so that one is
lead to
conclude
\begin{equation}
{\cal A}=-\lim_{a\rightarrow\infty}f(a,r)+\lim_{a\rightarrow
0}f(a,r).
\end{equation}
In these two limits the integral (\ref{inte}) is easily evaluated. It
gives
$-16\pi^2$ and $-8 \pi^2$,
respectively, so that ${\cal A}=8\pi^2$. This result is clearly
wrong, however,
because we
found $24\pi^2$ in the Eguchi-Hanson coordinates, see (\ref{res}).
The point is that the function $f(a,r)$ is not continuous in the
entire range
of
values of $a$: there is a jump at $a=r$, which is due to the presence
of the
Dirac string
in the Gibbons-Hawking coordinates. Indeed, one can check that
\begin{equation}
\lim_{a\rightarrow r^+}f(a,r)={112\over 27}\pi^2,\quad\quad
\lim_{a\rightarrow
r^-}
f(a,r)=-{320\over 27}\pi^2,
\end{equation}
so that
\begin{equation}
{\cal A}=16\pi^2+{112\over 27}\pi^2+{320\over
27}\pi^2-8\pi^2=24\pi^2,
\end{equation}
correctly. This calculation illustrates the technical complications,
that one
has to deal with when using in the Gibbons-Hawking coordinates. This
is useful
in view of the future calculations with multicenter metrics, where
the
Gibbons-Hawking coordinates are the only ones available.

To conclude, let me present some computation of the kind of
(\ref{res}).

In the case of the Taub-Nut metric with
\begin{equation}
e^a=\left\{{1\over 2}\sqrt{x+2m\over
x}dx,\sqrt{x(x+2m)}\sigma_x,\sqrt{x(x+2m)}\sigma_y,
2m\sqrt{x\over x+2m}\sigma_z\right\}
\end{equation}
($x>0$) the calculation gives
\begin{equation}
\omega^{(1)}_x=-96\cdot 16 \pi^2{x^2 m^2 \, dm\over
(x+2m)^5},\quad\quad
\int_0^\infty \omega^{(1)}_r=-16\pi^2.
\end{equation}

For the ${\bf CP}^2$ Fubini-Study metric with
\begin{equation}
e^a=\left\{{dr\over 1+\Lambda r^2},{r\sigma_x\over \sqrt{1+\Lambda
r^2}},
{r\sigma_y\over \sqrt{1+\Lambda r^2}},{r\sigma_z\over 1+\Lambda r^2}
\right\},
\end{equation}
one gets
\begin{equation}
\omega^{(1)}_r=-48 \pi^2{r^4\Lambda\, d\Lambda\over (1+\Lambda
r^2)^3},\quad
\quad \int_0^\infty
\omega^{(1)}_r=-24\pi^2.
\end{equation}

\end{document}